\documentclass[reqno,12pt,a4paper]{amsart}

\usepackage{mathrsfs}
\usepackage{amssymb}
\usepackage{amsfonts}
\usepackage{latexsym}
\usepackage{amsthm}

\usepackage{graphicx}
\def\lb{\label}

\newcommand{\er}[1]{\textrm{(\ref{#1})}}

\begin{document}


\renewcommand{\theequation}{\arabic{section}.\arabic{equation}}
\theoremstyle{plain}
\newtheorem{theorem}{\bf Theorem}[section]
\newtheorem{lemma}[theorem]{\bf Lemma}
\newtheorem{corollary}[theorem]{\bf Corollary}
\newtheorem{proposition}[theorem]{\bf Proposition}
\newtheorem{definition}[theorem]{\bf Definition}
\newtheorem{remark}[theorem]{\it Remark}

\def\a{\alpha}  \def\cA{{\mathcal A}}     \def\bA{{\bf A}}  \def\mA{{\mathscr A}}
\def\b{\beta}   \def\cB{{\mathcal B}}     \def\bB{{\bf B}}  \def\mB{{\mathscr B}}
\def\g{\gamma}  \def\cC{{\mathcal C}}     \def\bC{{\bf C}}  \def\mC{{\mathscr C}}
\def\G{\Gamma}  \def\cD{{\mathcal D}}     \def\bD{{\bf D}}  \def\mD{{\mathscr D}}
\def\d{\delta}  \def\cE{{\mathcal E}}     \def\bE{{\bf E}}  \def\mE{{\mathscr E}}
\def\D{\Delta}  \def\cF{{\mathcal F}}     \def\bF{{\bf F}}  \def\mF{{\mathscr F}}
\def\c{\chi}    \def\cG{{\mathcal G}}     \def\bG{{\bf G}}  \def\mG{{\mathscr G}}
\def\z{\zeta}   \def\cH{{\mathcal H}}     \def\bH{{\bf H}}  \def\mH{{\mathscr H}}
\def\e{\eta}    \def\cI{{\mathcal I}}     \def\bI{{\bf I}}  \def\mI{{\mathscr I}}
\def\p{\psi}    \def\cJ{{\mathcal J}}     \def\bJ{{\bf J}}  \def\mJ{{\mathscr J}}
\def\vT{\Theta} \def\cK{{\mathcal K}}     \def\bK{{\bf K}}  \def\mK{{\mathscr K}}
\def\k{\kappa}  \def\cL{{\mathcal L}}     \def\bL{{\bf L}}  \def\mL{{\mathscr L}}
\def\l{\lambda} \def\cM{{\mathcal M}}     \def\bM{{\bf M}}  \def\mM{{\mathscr M}}
\def\L{\Lambda} \def\cN{{\mathcal N}}     \def\bN{{\bf N}}  \def\mN{{\mathscr N}}
\def\m{\mu}     \def\cO{{\mathcal O}}     \def\bO{{\bf O}}  \def\mO{{\mathscr O}}
\def\n{\nu}     \def\cP{{\mathcal P}}     \def\bP{{\bf P}}  \def\mP{{\mathscr P}}
\def\r{\rho}    \def\cQ{{\mathcal Q}}     \def\bQ{{\bf Q}}  \def\mQ{{\mathscr Q}}
\def\s{\sigma}  \def\cR{{\mathcal R}}     \def\bR{{\bf R}}  \def\mR{{\mathscr R}}
\def\S{\Sigma}  \def\cS{{\mathcal S}}     \def\bS{{\bf S}}  \def\mS{{\mathscr S}}
\def\t{\tau}    \def\cT{{\mathcal T}}     \def\bT{{\bf T}}  \def\mT{{\mathscr T}}
\def\f{\phi}    \def\cU{{\mathcal U}}     \def\bU{{\bf U}}  \def\mU{{\mathscr U}}
\def\F{\Phi}    \def\cV{{\mathcal V}}     \def\bV{{\bf V}}  \def\mV{{\mathscr V}}
\def\P{\Psi}    \def\cW{{\mathcal W}}     \def\bW{{\bf W}}  \def\mW{{\mathscr W}}
\def\o{\omega}  \def\cX{{\mathcal X}}     \def\bX{{\bf X}}  \def\mX{{\mathscr X}}
\def\x{\xi}     \def\cY{{\mathcal Y}}     \def\bY{{\bf Y}}  \def\mY{{\mathscr Y}}
\def\X{\Xi}     \def\cZ{{\mathcal Z}}     \def\bZ{{\bf Z}}  \def\mZ{{\mathscr Z}}
\def\O{\Omega}
\def\th{\theta}

\newcommand{\gA}{\mathfrak{A}}
\newcommand{\gB}{\mathfrak{B}}
\newcommand{\gC}{\mathfrak{C}}
\newcommand{\gD}{\mathfrak{D}}
\newcommand{\gE}{\mathfrak{E}}
\newcommand{\gF}{\mathfrak{F}}
\newcommand{\gG}{\mathfrak{G}}
\newcommand{\gH}{\mathfrak{H}}
\newcommand{\gI}{\mathfrak{I}}
\newcommand{\gJ}{\mathfrak{J}}
\newcommand{\gK}{\mathfrak{K}}
\newcommand{\gL}{\mathfrak{L}}
\newcommand{\gM}{\mathfrak{M}}
\newcommand{\gN}{\mathfrak{N}}
\newcommand{\gO}{\mathfrak{O}}
\newcommand{\gP}{\mathfrak{P}}
\newcommand{\gQ}{\mathfrak{Q}}
\newcommand{\gR}{\mathfrak{R}}
\newcommand{\gS}{\mathfrak{S}}
\newcommand{\gT}{\mathfrak{T}}
\newcommand{\gU}{\mathfrak{U}}
\newcommand{\gV}{\mathfrak{V}}
\newcommand{\gW}{\mathfrak{W}}
\newcommand{\gX}{\mathfrak{X}}
\newcommand{\gY}{\mathfrak{Y}}
\newcommand{\gZ}{\mathfrak{Z}}

\def\ve{\varepsilon}   \def\vt{\vartheta}    \def\vp{\varphi}    \def\vk{\varkappa}

\def\Z{{\mathbb Z}}    \def\R{{\mathbb R}}   \def\C{{\mathbb C}}    \def\K{{\mathbb K}}
\def\T{{\mathbb T}}    \def\N{{\mathbb N}}   \def\dD{{\mathbb D}}


\def\la{\leftarrow}              \def\ra{\rightarrow}            \def\Ra{\Rightarrow}
\def\ua{\uparrow}                \def\da{\downarrow}
\def\lra{\leftrightarrow}        \def\Lra{\Leftrightarrow}


\def\lt{\biggl}                  \def\rt{\biggr}
\def\ol{\overline}               \def\wt{\widetilde}
\def\no{\noindent}


\let\ge\geqslant                 \let\le\leqslant
\def\lan{\langle}                \def\ran{\rangle}
\def\/{\over}                    \def\iy{\infty}
\def\sm{\setminus}               \def\es{\emptyset}
\def\ss{\subset}                 \def\ts{\times}
\def\pa{\partial}                \def\os{\oplus}
\def\om{\ominus}                 \def\ev{\equiv}
\def\iint{\int\!\!\!\int}        \def\iintt{\mathop{\int\!\!\int\!\!\dots\!\!\int}\limits}
\def\el2{\ell^{\,2}}             \def\1{1\!\!1}
\def\sh{\sharp}
\def\wh{\widehat}
\def\bs{\backslash}

\def\sh{\mathop{\mathrm{sh}}\nolimits}
\def\Area{\mathop{\mathrm{Area}}\nolimits}
\def\arg{\mathop{\mathrm{arg}}\nolimits}
\def\const{\mathop{\mathrm{const}}\nolimits}
\def\det{\mathop{\mathrm{det}}\nolimits}
\def\diag{\mathop{\mathrm{diag}}\nolimits}
\def\diam{\mathop{\mathrm{diam}}\nolimits}
\def\dim{\mathop{\mathrm{dim}}\nolimits}
\def\dist{\mathop{\mathrm{dist}}\nolimits}
\def\Im{\mathop{\mathrm{Im}}\nolimits}
\def\Iso{\mathop{\mathrm{Iso}}\nolimits}
\def\Ker{\mathop{\mathrm{Ker}}\nolimits}
\def\Lip{\mathop{\mathrm{Lip}}\nolimits}
\def\rank{\mathop{\mathrm{rank}}\limits}
\def\Ran{\mathop{\mathrm{Ran}}\nolimits}
\def\Re{\mathop{\mathrm{Re}}\nolimits}
\def\Res{\mathop{\mathrm{Res}}\nolimits}
\def\res{\mathop{\mathrm{res}}\limits}
\def\sign{\mathop{\mathrm{sign}}\nolimits}
\def\span{\mathop{\mathrm{span}}\nolimits}
\def\supp{\mathop{\mathrm{supp}}\nolimits}
\def\Tr{\mathop{\mathrm{Tr}}\nolimits}
\def\BBox{\hspace{1mm}\vrule height6pt width5.5pt depth0pt \hspace{6pt}}
\def\as{\text{as}}
\def\all{\text{all}}
\def\where{\text{where}}
\def\Dom{\mathop{\mathrm{Dom}}\nolimits}


\newcommand\nh[2]{\widehat{#1}\vphantom{#1}^{(#2)}}
\def\dia{\diamond}

\def\Oplus{\bigoplus\nolimits}



\def\qqq{\qquad}
\def\qq{\quad}
\let\ge\geqslant
\let\le\leqslant
\let\geq\geqslant
\let\leq\leqslant
\newcommand{\ca}{\begin{cases}}
\newcommand{\ac}{\end{cases}}
\newcommand{\ma}{\begin{pmatrix}}
\newcommand{\am}{\end{pmatrix}}
\renewcommand{\[}{\begin{equation}}
\renewcommand{\]}{\end{equation}}
\def\eq{\begin{equation}}
\def\qe{\end{equation}}
\def\[{\begin{equation}}
\def\bu{\bullet}

\title[{Spectral asymptotics for the third order operator}]
{Spectral asymptotics for the third order operator with
periodic coefficients}

\date{\today}
\author[Andrey Badanin]{Andrey Badanin}
\address{A. Badanin: Northern (Arctic) Federal University,
Russia, e-mail: an.badanin@gmail.com}
\author[Evgeny Korotyaev]{Evgeny Korotyaev}
\address{E. Korotyaev: St.-Petersburg State
University, Russia, e-mail: korotyaev@gmail.com}

\subjclass{47E05, 34E10}
\keywords{third order operator with periodic coefficients,
spectral asymptotics}

\begin{abstract}
We consider the self-adjoint third order operator  with
1-periodic coefficients on the real line.
The spectrum of the operator is absolutely continuous and covers the
real line.  We determine the high energy asymptotics of the
periodic, anti-periodic eigenvalues and of the branch points of the
Lyapunov function. Furthermore, in the case of small coefficients we
show that either whole  spectrum has multiplicity one or the spectrum
has multiplicity one except for a small spectral nonempty interval
with multiplicity three. In the last case the asymptotics of this
small interval is determined.

\end{abstract}

\maketitle


\section {Introduction and main results}
\setcounter{equation}{0}

We consider the self-adjoint operator $H$ acting in $L^2(\R)$ and
given by
\[
\lb{Hpq} H=i\pa^3+ip\pa+i\pa p+q
\]
where the  real  1-periodic coefficients $p,q$ belong to the space
$L^1(\T),\T=\R/\Z$, equipped with the norm $\|f\|=\int_0^1|f(s)|ds$.
Without loss of generality we assume that
\[
\lb{0} \int_0^1 q(t)dt=0.
\]

A great number of papers is devoted to the inverse spectral theory
for the Schr\"odinger operator with periodic potential: Dubrovin
\cite{D}, Garnett--Trubowitz \cite{GT}, Its--Matveev \cite{IM},
Kargaev--Korotyaev \cite{KK}, Marchenko--Ostrovski \cite{MO},
Novikov \cite{No} etc. Note that Korotyaev \cite{K3} extended the
results of \cite{MO}, \cite{GT}, \cite{KK}, for the case $-y''+qy$
to the case of periodic distributions, i.e., $-y''+q'y$ on $L^2(\R)$,
where periodic $q\in L_{loc}^2(\R)$.

The results for the operator $H$ are used in the
integration of the {\it bad Boussinesq equation}, given by
\[
\lb{Be}
\ddot p={1\/3}\pa^2\Bigl(\pa^2 p+{4}p^2\Bigr),\qq\dot p=\pa q,
\]
on the circle,
see \cite{McK} and references therein.
Here $ \dot u$ (or $\pa u$) means the derivatives of the function
$u$ with respect to the time (or space) variable. It is equivalent to the
Lax equation $\dot H=HK-KH$ where
$K=-\pa^2+{4\/3}p$.

The inverse scattering
theory for the self-adjoint third order operator $i\pa^3+ip\pa+i\pa p+q$
with decreasing coefficients was
developed in \cite{DTT}.
McKean \cite{McK}
obtained the numerous results in the inverse spectral theory for the
non-self-adjoint operator $H_*=\pa^3+p\pa+\pa p+q$ on the real line
with smooth and sufficiently small $p$ and $q$.
Results for  non-self-adjoint operator $H_*$ was
applied for integration of  the
{\it good Boussinesq equation}.

It is known much less about the self-adjoint third order operators
with periodic coefficients.
Even the direct problem is not well developed.

Results  about the spectrum of  the  higher order differential operators with smooth
periodic coefficients are given in the book \cite{DS}, see also McGarvey's paper \cite{McG}.
The case of the even order differential operators with non-smooth
coefficients are given in \cite{BK3}.

The operator $H$ was considered by the authors in \cite{BK4}.
In order to describe our main results we  recall needed results from \cite{BK4}.
We consider the differential equation
\[
\lb{1b} i y'''+ipy'+i(py)'+q y=\l y,\ \ \ (t,\l)\in\R\ts\C.
\]
Introduce the $3\ts 3$ matrix-valued function $M(t,\l)$ by
\[
\lb{deM}
M(t,\l)=\ma\vp_1&\vp_2&\vp_3\\
\vp_1'&\vp_2'&\vp_3'\\
\vp_1''+p\vp_1&\vp_2''+p\vp_2&\vp_3''+p\vp_3\am(t,\l),\qq
(t,\l)\in\R\ts\C,
\]
where $\vp_1, \vp_2, \vp_3$ are the fundamental solutions of equation
\er{1b} satisfying the conditions
\[
\lb{ic} M(0,\l) =\1_3,\qqq \forall \ \l\in\C.
\]
Henceforth $\1_3$ is the $3\ts 3$ identity matrix. We define
{\it the modified monodromy matrix} by $M(1,\l)$, below it
 is called shortly {\it the monodromy matrix}.
The matrix-valued function $M(1,\cdot)$ is entire and its
characteristic polynomial $D$ is given by
\[
\lb{1c} D(\t,\l)=\det(M(1,\l)-\t \1_{3}),\qq (\t,\l)\in\C^2.
\]
An eigenvalue  of $M(1,\l)$ is called a {\it multiplier}, it is a
zero of the algebraic equation $D(\cdot,\l)=0$. If $\t(\l)$ is a
multiplier for some $\l\in \C$, then $\ol\t(\ol\l)^{-1}$ is also a
multiplier. Each $M(1,\l), \l\in\C$, has exactly $3$ (counting with
multiplicities) multipliers $\t_j(\l),j=1,2,3$, which  satisfy
\[
\lb{lom} \t_j(\l)=e^{iz\o^{j-1}}(1+O(|z|^{-1}))\qq\text{as} \qq
|\l|\to\iy.
\]
Henceforth we put
$$
z=\l^{1\/3},\qqq\arg\l\in\Big(-{\pi\/2},{3\pi\/2}\Big], \qqq\arg
z\in\Big(-{\pi\/6},{\pi\/2}\Big], \qqq\o=e^{i{2\pi\/3}}.
$$

The operator $H$ is self-adjoint on the domain
\[
\lb{cDH} \Dom(H)=\rt\{f\in L^2(\R):i(f''+pf)'+ipf'+qf\in L^2(\R),
f'',(f''+pf)'\in L_{loc}^1(\R)\rt\}.
\]
The spectrum of $H$ is absolutely continuous and satisfies
$$
\s(H)=\rt\{\l\in\R:|\t_j(\l)|=1,\ \text{some}\
j=1,2,3\rt\}=\gS_1\cup \gS_3=\R,\qqq \gS_2=\es,
$$
where $\gS_j$ is the part of the spectrum of $H$ having the
multiplicity $j=1,2,3$. The spectrum of $H$ has multiplicity 1 at
high energy. Note that the spectrum of the even order operator
with periodic coefficients has multiplicity 2 at high energy,
see \cite{BK3}.
Recall that the spectrum of  the Hill operator
$-\pa^2+q$ on the real line is absolutely continuous and consists of
spectral bands separated by gaps. In the case of the "generic"
potential all gaps in the spectrum are open, see \cite{MO},
\cite{K3}. In the case of third order operators with periodic
coefficients there are
no gaps in the spectrum for all $p,q$.

The coefficients of the polynomial $D(\cdot,\l)$ are entire
functions of $\l$. Due to \er{lom} there exist exactly three roots
$\t_1(\cdot), \t_2(\cdot)$ and $\t_3(\cdot)$, which
constitute three distinct branches of some analytic function
$\t(\cdot)$ that has only algebraic singularities in $\C$, see,
e.g.,  Ch.~8 in \cite{Fo}. Thus the function $\t(\cdot)$ is
analytic on some 3-sheeted Riemann surface $\cR$.
There are only a finite number of the algebraic singularities  in
any bounded domain. In order to describe these points we introduce
the {\it discriminant} $\r(\l),\l\in\C$, of the polynomial
$D(\cdot,\l)$ by
\[
\lb{r}
\r=(\t_1-\t_2)^2(\t_1-\t_3)^2(\t_2-\t_3)^2.
\]
The function $\r=\r(\l)$ is entire and real on $\R$. Due to McKean
\cite{McK} a zero of $\r$ is called a {\it ramification} and it is
the branch point of the corresponding  Riemann surface $\cR$.
Ramification is a geometric term used for 'branching out', in the
way that the square root function, for complex numbers, can be seen
to have two branches differing in sign. We also use it from the
opposite perspective (branches coming together) as when a covering
map degenerates at a point of a space, with some collapsing together
of the fibers of the mapping. In the case of the non-self-adjoint
operator $H_*$
the ramifications are invariant with
respect to the Boussinesq flow and they can consider as the spectral
data, see \cite{McK}.

If $p=q=0$, then the function $\r$ and its zeros $r_{n}^{0,\pm}$
have the form
\[
\lb{ro0} \r_0=64\sinh^2{\sqrt 3 z\/2}\sinh^2{\sqrt 3 \o
z\/2}\sinh^2{\sqrt 3 \o^2z\/2},\qqq
r_{n}^{0,+}=r_{n}^{0,-}=i\rt({2\pi n\/\sqrt3}\rt)^3,\ \ n\in\Z.
\]
We formulate our first results about the zeros $r_n^\pm,n\in\Z$, of
the function $\r$ (the ramifications).

\begin{theorem}
\lb{Thr} i) The function $\r$ is entire, real on $\R$, and satisfies
\[
\lb{rtr} \r(\l)= |T(\l)|^4-8\Re T^3(\l)+18|T(\l)|^2-27, \qq \forall\qq
\l\in\R,
\]
where
\[
\lb{defT} T(\cdot)=\Tr M(1,\cdot).
\]

ii) Let $\gS_3$ be the part of the spectrum of $H$ having the
multiplicity 3. Then
\[
\lb{sro} \gS_3=\{\l\in\R:|\t_j(\l)|=1,\ \forall\
j=1,2,3\}=\{\l\in\R:\r(\l)\le 0\}.
\]
Moreover, there exists only a finite number $(\ge 0)$ of the bounded
spectral bands with the spectrum of multiplicity $3$.

iii) The ramifications $r_{n}^\pm$ as $n\to+\iy$ satisfy:
\[
\begin{aligned}
\lb{are} r_{n}^\pm=\ol{r_{-n}^\mp}=r_{n}^{0,\pm}-i{4\pi n\/\sqrt3}\Big(\wh p_0\mp
|\wh p_n|+O(w_n)+O(n^{-1})\Big),\\
\wh p_n=\int_0^1p(t)e^{-i2\pi nt}dt,\qq n\in\Z,
\end{aligned}
\]
for some  sequence $w_n, n\in \N$,
such that $\sum _{n\ge 1}{|w_n|^2\/n}<\iy$.
\end{theorem}

\no {\bf Remark.} i) Identities \er{sro} show that the endpoints
of every spectral interval with the spectrum
of multiplicity 3 are the ramifications.

ii) In this paper we analyze the spectrum at high energy using the
multipliers. The functions
$\D_j={1\/2}(\t_j+\t_j^{-1}),j=1,2,3, $ are the branches of the
standard Lyapunov function $\D(\l)$ analytic on
a 3-sheeted Riemann surface. This function
is more convenient for analysis of the spectrum at finite
energy. Note that identity \er{sro} implies:
$$
\gS_3=\{\l\in\R:\D_j(\l)\in[-1,1],\ \forall\ j=1,2,3\}.
$$
The graph of a typical Lyapunov function
and the spectrum $\gS_3$ are shown by
Fig.~\ref{fig2}.

\begin{figure}
\tiny
\unitlength 1mm 
\linethickness{0.4pt}
\ifx\plotpoint\undefined\newsavebox{\plotpoint}\fi 
\begin{picture}(116.85,116.9)(0,0)
\put(6.25,71.9){\line(0,-1){58.4}}
\put(4.65,22.9){\line(1,0){107.6}}
\put(5.05,42.5){\line(1,0){105.4}}
\put(4.65,62.5){\line(1,0){105.4}}
\put(111.05,38.9){\makebox(0,0)[cc]{$\l$}}
\put(10.25,68.3){\makebox(0,0)[cc]{$\D(\l)$}}
\put(2.85,39.5){\makebox(0,0)[cc]{$0$}}
\put(3.05,19.1){\makebox(0,0)[cc]{$-1$}}
\put(2.85,59.3){\makebox(0,0)[cc]{$1$}}
\thicklines
\qbezier(9.25,24.9)(20.65,20.2)(26.45,27.9)
\qbezier(20.85,42.5)(30.45,33.5)(26.45,27.9)
\qbezier(23.45,57.5)(15.15,49)(20.85,42.5)
\qbezier(38.25,59.9)(32.05,65.9)(23.45,57.5)
\qbezier(38.25,59.9)(42.85,56.1)(37.85,28.3)
\qbezier(37.85,28.3)(36.05,19)(31.05,26.9)
\qbezier(31.05,26.9)(29.65,28.3)(34.65,29.9)
\qbezier(34.65,29.9)(54.35,37.9)(57.65,43)
\qbezier(57.65,43)(71.35,62.4)(57.45,62.3)
\qbezier(57.65,43)(45.75,62.4)(57.45,62.3)
\qbezier(80.65,29.9)(60.95,37.9)(57.65,43)
\qbezier(84.25,26.7)(85.65,28.3)(80.65,29.9)
\qbezier(77.45,28.3)(79.25,19)(84.25,26.7)
\qbezier(77.05,60.1)(72.45,56.1)(77.45,28.3)
\qbezier(77.05,60.1)(82.85,65.9)(91.45,57.5)
\qbezier(91.45,57.5)(100.75,49)(94.05,42.5)
\qbezier(94.05,42.5)(84.45,33.5)(89.95,28.1)
\qbezier(107.05,24.9)(97.65,20.2)(89.95,28.1)
\put(6.25,116.9){\line(0,-1){38.4}}
\put(5.05,87.5){\line(1,0){105.4}}
\put(109,83){\makebox(0,0)[cc]{$\l$}}
\put(8.25,83.5){\makebox(0,0)[cc]{$0$}}
\put(12.25,116.25){\makebox(0,0)[cc]{$\r(\l)$}}
\put(20.65,90.25){\makebox(0,0)[cc]{$r_0^-$}}
\put(27,90.25){\makebox(0,0)[cc]{$r_{-1}^+$}}
\put(34,90.25){\makebox(0,0)[cc]{$r_{-1}^-$}}
\put(39.5,90.25){\makebox(0,0)[cc]{$r_{-2}^+$}}
\put(53,90.25){\makebox(0,0)[cc]{$r_{-2}^-$}}
\put(97,90.25){\makebox(0,0)[cc]{$r_0^+$}}
\put(90,90.25){\makebox(0,0)[cc]{$r_{1}^-$}}
\put(84,90.25){\makebox(0,0)[cc]{$r_{1}^+$}}
\put(76,90.25){\makebox(0,0)[cc]{$r_{2}^-$}}
\put(64,90.25){\makebox(0,0)[cc]{$r_{2}^+$}}
\thicklines
\qbezier(18.5,87.5)(14.875,92.625)(9.75,113.25)
\qbezier(18.5,87.5)(20.75,84.5)(28,87.5)
\qbezier(28,87.5)(30.5,88.75)(31,87.5)
\qbezier(31,87.5)(36.875,83.5)(40.75,87.5)
\qbezier(40.75,87.5)(47.5,95.375)(51.25,87.5)
\qbezier(51.25,87.5)(55.625,76.375)(58,75.75)
\qbezier(64.75,87.5)(60.375,76.375)(58,75.75)
\qbezier(75.25,87.75)(68.5,95.375)(64.75,87.5)
\qbezier(84.5,87.5)(79.125,83.5)(75.25,87.5)
\qbezier(88,87.5)(85.5,88.75)(84.5,87.5)
\qbezier(97.5,87.5)(95.25,84.5)(88,87.5)
\qbezier(97.5,87.5)(101.125,92.625)(106.25,113.25)
\thinlines
\put(4.25,3.5){\line(1,0){112.6}}
\put(100,5.75){\makebox(0,0)[cc]{$\gS_3$}}
\linethickness{4pt}
\put(18.65,3.9){\line(1,0){9.2}}
\put(97.05,3.9){\line(-1,0){9.2}}
\put(31.25,3.9){\line(1,0){9.8}}
\put(84.45,3.9){\line(-1,0){9.8}}
\put(51.45,3.9){\line(1,0){8.2}}
\put(64.25,3.9){\line(-1,0){8.2}}
\thinlines
\multiput(18.43,87.43)(.002404,-.995192){85}{{\rule{.4pt}{.4pt}}}
\multiput(97.18,87.18)(-.002404,-.995192){85}{{\rule{.4pt}{.4pt}}}
\multiput(27.68,87.68)(0,-.992788){85}{{\rule{.4pt}{.4pt}}}
\multiput(87.93,87.43)(0,-.992788){85}{{\rule{.4pt}{.4pt}}}
\multiput(31.18,87.18)(0,-.990385){85}{{\rule{.4pt}{.4pt}}}
\multiput(84.43,86.93)(0,-.990385){85}{{\rule{.4pt}{.4pt}}}
\multiput(40.68,87.68)(0,-.992788){85}{{\rule{.4pt}{.4pt}}}
\multiput(74.93,87.43)(0,-.992788){85}{{\rule{.4pt}{.4pt}}}
\multiput(51.18,87.43)(.004808,-.995192){85}{{\rule{.4pt}{.4pt}}}
\multiput(64.43,87.18)(-.004808,-.995192){85}{{\rule{.4pt}{.4pt}}}
\end{picture}
\lb{fig2}
\caption{\footnotesize
The function $\r$, the Lyapunov function and the spectrum
$\gS_3$}
\end{figure}
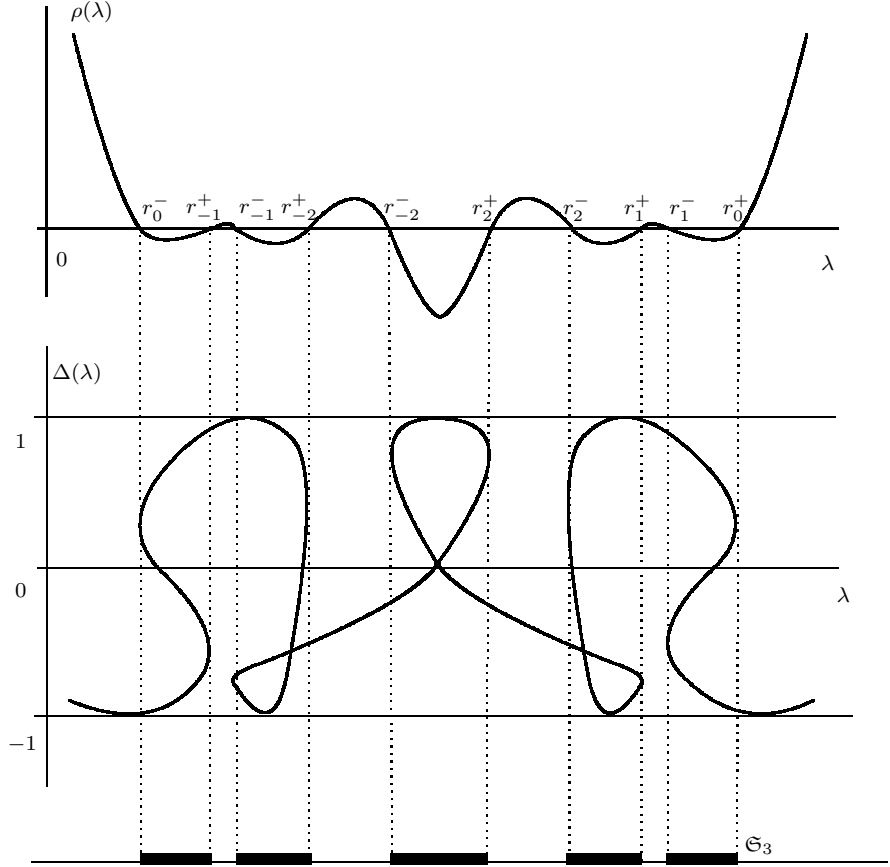

Let $(\l_{2n})_{n\in\Z}$ be the sequence of eigenvalues of
equation \er{1b} with the 1-periodic boundary condition
$y(x+1)=y(x), x\in \R$. Let $(\l_{2n+1})_{n\in\Z}$ be the sequence
of eigenvalues of the equation \er{1b} with the anti-periodic
boundary condition $y(x+1)=-y(x), x\in \R$. They are labeling
(counted with multiplicity) by
\[
\begin{aligned}
\lb{l2p}
...\le\l_{-4}\le\l_{-2}\le\l_0\le\l_2\le\l_4\le..., \qqq\qqq
\text{the periodic eigenvalues},\\
...\le\l_{-3}\le\l_{-1}\le\l_1\le\l_3\le..., \qqq
\text{the anti-periodic eigenvalues}.
\end{aligned}
\]
If $p=q=0$, then the periodic and antiperiodic eigenvalues are given
by $\l_n^0=(\pi n)^3,n\in\Z$.

\begin{theorem}\lb{Per}

i) The periodic and antiperiodic eigenvalues satisfy
\[
\lb{apa} \l_n=(\pi n)^3-2\wh p_0\pi n+{o(1)\/n}\qq\as\qq
n\to\pm\iy.
\]

ii) The entire function $T=\Tr M(1,\cdot)$ and the spectrum $\gS_3$
of the multiplicity three are recovered by the periodic spectrum
plus one antiperiodic eigenvalue or by the antiperiodic spectrum
plus one periodic eigenvalue.

\end{theorem}

\no {\bf Remark.} i) Asymptotics \er{are} of the ramification is
sharp, since it is determined in terms of Fourier coefficients of
$p,q$. Unfortunately, the asymptotics of the periodic and
antiperiodic eigenvalues in \er{apa} is not sharp.

ii) Knowing the function $T$ and using identities \er{rtr}, \er{cM}
we recover the functions $\r(\l), D(\t,\l)$ and all multipliers.

\medskip

We consider the operator $H_\ve$ acting in $L^2(\R)$ on the domain
\er{cDH} and given by
\[
\lb{Hve} H_\ve=i\pa^3+\ve (ip\pa+i\pa p+q)
\]
where $\ve\in\R$ is a small coupling constant. We have the following
result.

\begin{theorem}
\lb{1.3} Let $p\in L^1(\T)$ satisfy $\int_0^1 p(t)dt=0$. Then there
exist two functions $r^\pm(\ve)$, real analytic in the disk
$\{|\ve|<c\}\ss \C$ for some $c>0$, such that $r^\pm(0)=0$ and they
satisfy
\[
\lb{l0} r^+(\ve)-r^-(\ve)=4h^{3\/2}\ve^3+O(\ve^4) \qq\as\qq\ve\to 0,
\]
\[
\gS_3=\ca  (r^-(\ve),r^+(\ve)) \ {\rm or }   \  (r^+(\ve),r^-(\ve)) \  & if \ h>0\\
               \es       & if \ h<0\ac,
\]
where
\[
\lb{F1} h={2\/3}\sum_{n\ge 1}\Bigl({|\wh p_n|^2\/(2\pi n)^2} -{3|\wh
q_n|^2\/(2\pi n)^4}\Bigr).
\]
\end{theorem}

{\bf Remark.} i) The functions $r^\pm(\ve)$ are the (nearest to $\l=0$)
zeros of the function $\r(\l,\ve)$.

ii) Let $\ve>0$ be small enough. If $h>0$, then $r^\pm(\ve)$  are
real and there is a band of the spectrum of the multiplicity
three. If $h<0$, then $r^\pm(\ve)$  are
non-real and there is no a band of the spectrum of the multiplicity
three.

iii) The proof of the theorem is based on the analysis of the
monodromy matrix as $\ve\to 0$, and  identities \er{rtr}, \er{sro}.
We determine the asymptotics of
$r^\pm(\ve)$ in the form $r^\pm(\ve)=r(\ve)\pm
2h^{3\/2}\ve^3+O(\ve^4)$ as $\ve\to 0$ for some function $r$.
This gives the asymptotics \er{l0}. In the proof we use the some
technique developed for  fourth order operator with small $p,q$, see
\cite{BK1, BK2}.

We describe now the results for vector differential equations and
higher order differential equations. We begin with the vector case,
where more deep results are obtained. The inverse problem for
vector-valued Sturm-Liouville operators on the unit interval with
Dirichlet boundary conditions, including characterization, was
solved by Chelkak--Korotyaev \cite{CK1},  \cite{CK2}. We mention
that uniqueness for inverse problems for systems on finite intervals
was studied in \cite{Ma}. The periodic case is more complicated and
a lot of papers are devoted only to the direct problem for periodic
systems: Carlson \cite{Ca1}, \cite{Ca2},  Gelfand--Lidskii
\cite{GL}, Gesztesy and coauthors \cite{CHGL}, \cite{CG}, Korotyaev
and coauthors \cite{CK}, \cite{BBK}, \cite{K1}, \cite{K2}, etc.
The discrete periodic systems  were studied in
\cite{KKu1}, \cite{KKu2}.
We describe results for the first and second order operators with the
periodic $N\ts N$ matrix-valued potential from \cite{CK}, \cite{K1},
\cite{K2}, which are important for us. In fact the direct problem is
consisted from two steps:

First step:

(1) the Lyapunov function on some Riemann surface is constructed and
described,

(2) sharp asymptotics of periodic eigenvalues and ramifications of
the Lyapunov function are determined,

(3) multiplicity of the spectrum, endpoints of gaps are the periodic
or antiperiodic eigenvalues or the ramifications of the Lyapunov
function are determined.

The  second step is more complicated:

(4) the conformal mapping with real part given by the integrated
density of states and imaginary part given by the Lyapunov exponent
is constructed and the main properties are obtained,

(5) trace formulas (similar to the case of the Hill operators) are
determined,

(6) global estimates of gap lengths in terms of $L^2$--norm of
potentials are obtained.

 Spectral theory for higher order operators with decreasing coefficients
  is well developed, see \cite{BDT} and the references therein.
Numerous papers are devoted to the fourth order operators on bounded
interval: \cite{B}, \cite{CPS}, \cite{McL}, \cite{S} etc. The third
order operator on the bounded interval was considered by Amour
\cite{A1}, \cite{A2}.

Even ($\ge 4$) order operators with periodic coefficients considered
in the papers: Badanin--Korotyaev \cite{BK1} -- \cite{BK3},
Papanicolaou \cite{P1,P2},
Mikhailets--Molyboga \cite{MM1,MM2}, Tka\-chen\-ko \cite{Tk}, see also
references therein. The spectral analysis of the higher ($\ge 3$)
order operators with periodic coefficients is more difficult, than
the analysis of the first and second order systems with periodic
matrix-valued potentials. The main reason is that the monodromy
matrix for the first and second order systems has asymptotics in
terms of $\cos $ and  $\sin $ bounded  on the real line. The
asymptotics of the monodromy matrix for higher order operators has
additional components in terms  of $\cosh $ and  $\sinh $ unbounded
on the real line.

The $2N$-order ($N\ge 2$) operator with periodic coefficients was
considered in \cite{BK3} (the case $N=2$ see also in \cite{BK1},
\cite{BK2}) and only the first step was done.
 The conformal
mapping for the higher order operator, which is important for the
spectral analysis, is not still constructed and there are no gap
length estimates in terms of the norms of potentials.

The plan of the paper is as follows.
In Sect. 2 we describe the basic properties of the monodromy matrix $\cM$.
In Sect. 3 we consider the function $\r$ and
the Riemann surface of the multipliers at high energy
and prove Theorem \ref{Thr} i),ii).
In Sect. 4 we determine asymptotics of the ramifications
and prove Theorem \ref{Thr} iii).
In Sect. 5 we determine asymptotics of the periodic and antiperiodic spectrum
and prove Theorem \ref{Per}.
In Sect. 6 we consider the case of the small coefficients
and prove Theorem \ref{1.3}.
The technical proof of the asymptotics of the trace
of the monodromy matrix $\cM$ is placed in Appendix.

\section {Monodromy matrix}
\setcounter{equation}{0}

Consider the unperturbed equation $iy'''=\l y$. It has the solutions
$e^{iz\o^{j-1}t}$ and the multipliers have the form
$e^{i\o^{j-1}z},j=1,2,3,$
here and below
\[
\lb{S}
\o=e^{i{2\pi\/3}},\qq z=x+iy=\l^{1\/3}\in S,\qq
\arg\l\in\Bigl(-{\pi\/2},{3\pi\/2}\Bigr], \qq S=\{z\in \C:\arg
z\in\Bigl(-{\pi\/6},{\pi\/2}\Bigr]\},
\]
and {\it on the boundary of $S$ we identify each point
$z=re^{i{\pi\/2}},r>0$,  with the point $re^{-i{\pi\/6}}$}. The
trace $\Tr M_0$ of the unperturbed monodromy matrix $M_0$ is an
entire function in $\l$ given by
\[
\lb{t01} T_0=\Tr M_0=e^{iz}+e^{i\o z}+e^{i\o^2 z}.
\]

Consider the perturbed equation \er{1b}.
The matrix-valued function $M(t,\l)$, given by \er{deM}, satisfies
\[
\lb{me1}
M'- P(\l)M= Q(t)M,\qq M(0,\l)=\1_3,\qq\all\qq (t,\l)\in\R\ts\C,
\]
where the $3\ts 3$ matrices $ P$ and $ Q$ have the form
\[
\lb{mu}
 P=\ma 0&1&0\\0 &0&1\\-i\l&0&0\am,\qq
 Q=\ma 0&0&0\\-p&0&0\\iq&-p&0\am.
\]
The matrix-valued function $M(1,\cdot)$
is entire and satisfies
\[
\lb{idM}
\det M(1,\l)=1,\qq
M^*(1,\ol\l)JM(1,\l)=J\qq \  \qq\where\ \
J=\ma
0&0&i\\
0&-i&0\\
i&0&0
\am
\]
for all $\l\in\C$, see \cite{BK4}. This identity is an odd-dimensional analog
of the symplectic property of the monodromy matrix for
the even order operator \cite{BK3}. Moreover,
\[
\lb{cM} D(\t,\l)=-\t^3+\t^2T(\l)-\t\ol
T(\ol\l)+1,\qq\all\qq(\t,\l)\in\C^2.
\]

Introduce the simple transformation
\[
\lb{js1}
\cM= \mU^{-1}M \mU,
\qqq
\mU^{-1} P\mU=iz\O,\qqq
\cQ=\mU^{-1} Q\mU={1\/3iz}\Bigl(p Q_1+{q\/z} Q_2\Bigr)
\]
where $\mU=\cZ U$ and
\[
\lb{dcU}
\O=\ma1&0&0\\ 0 &\o &0 \\
0&0&\o^2\am,\qq
\cZ
=\ma1&0&0\\ 0 &iz &0 \\
0&0&(iz)^2\am,\qq
U={1\/\sqrt3}\ma1&1&1\\ 1 &\o &\o^2 \\
1&\o^{2}&\o\am=(U^*)^{-1},
\]
\[
\lb{cK12}
Q_1=\ma -2&\o^2&\o\\
1&-2\o^2&\o\\
1&\o^2&-2\o\am,
\qq
 Q_2
=\ma 1& 1& 1\\\o&\o&\o\\\o^2&\o^2&\o^2\am.
\]
Using this transformation we rewrite the problem \er{me1} in the form
\[
\lb{me2}
\cM'-iz\O\cM=\cQ(t,\l)\cM,\qqq \cM(0,\l)=\1_3.
\]

Identities \er{js1}, \er{cK12} show that the matrix
$\cQ$ in the right hand side
of equation \er{me2} satisfies $\cQ(t,\l)=O((|p(t)|+|q(t)|)|z|^{-1})$ as $|\l|\to\iy$,
while the corresponding coefficient $Q$ in \er{me1}
is only bounded. It is a crucial point for our analysis.

In \cite{BK4} we proved that the monodromy matrix $\cM$ and its trace
$T(\l)=\Tr\cM(1,\l)$ satisfy
\[
\lb{bk41}
|\cM(1,\l)-e^{iz\O}|\le {\vk\/|z|}e^{z_0+\vk},\qq  \all\qq  |\l|\ge 1,
\]
\[
\lb{bk42}
\begin{aligned}
|T(\l)|\le 3e^{z_0+\vk},\qq\all\qq\l\in\C,\\
|T(\l)-T_0(\l)|\le {3\vk\/|z|}e^{z_0+\vk},\qq  \all\qq|\l|\ge 1
\end{aligned}
\]
where
\[
\lb{kz0}
\vk=\|p\|+\|q\|,\qq
z_0=\max_{j=0,1,2}\Re(iz\o^j)=\Re(iz\o^2)={y+\sqrt3x\/2},
\qq {|z|\/2}\le z_0\le|z|,
\]
and $z=x+iy$. Henceforth a matrix $A$ has the norm given by
\[
\lb{mnorm}
|A|=\max\{\sqrt{E}\ge0:E\ \text{is an eigenvalue of the matrix}\ A^*A\}.
\]

Below we need sharper estimates of the monodromy matrix.
Equation \er{me2} and the standard arguments  provide that the
function $\cM(t,\l)$ satisfies the integral equation
$$
\cM(t,\l)=e^{izt\O}+\int_0^te^{iz(t-s)\O}\cQ(s,\l)\cM(s,\l)ds,\qqq
\all\qq
(t,\l)\in\R\ts(\C\sm\{0\}),
$$
where the matrix $\O$ is given by \er{dcU}.
The standard iterations yield
\[
\lb{eNj}
\cM(t,\l)=\sum_{0}^\iy\cM_n(t,\l),\qqq \cM_0(t,\l)=e^{izt\O}
\]
where
\[
\lb{cNn}
\qq
\cM_n(t,\l)=\int_0^te^{iz(t-s)\O}\cQ(s,\l)\cM_{n-1}(s,\l)ds,
\qq n\in\N.
\]

\begin{lemma}
\lb{MM}
The series \er{eNj} converges absolutely and uniformly on any compact in
$\C\sm\{0\}$.
The matrix-valued function $\cM(1,\cdot)$ is analytic
in $\C\sm\{0\}$ and satisfies
\[
\lb{eN}
|\cM(1,\l)|\le e^{z_0+\vk},\qq
|\cM(1,\l)-\sum_0^{N-1}\cM_n(1,\l)|\le{e^{z_0+\vk}\/|z|^{N}},
\qq \all \qq N\in\N,\qq|\l|\ge 1.
\]
\end{lemma}

\no {\bf Proof.}  Identities \er{cNn} give
$$
\cM_n(t,\l)=\int\limits_{0< t_1<...< t_n<t_{n+1}=t}\prod\limits_{k=1}^{n}
\Bigl(e^{iz(t_{k+1}-t_k)\O}\cQ(t_k)\Bigr)\cM_0(t_1,\l)dt_1dt_2...dt_n,
$$
the factors are ordering from right to left.
Using estimates $|e^{izt\O}|\le e^{z_0t}$ we obtain
\[
\lb{eN1}
|\cM_n(t,\l)|\le{e^{z_0t}\/n!}\lt(\int_0^t|\cQ(s)|ds\rt)^n,
\qqq \all \qq (n,t,\l)\in\N\ts\R_+\ts(\C\sm\{0\}).
\]
These estimates show that for each fixed $t\in\R_+$ the formal
series \er{eNj} converges absolutely and uniformly on any compact in
$\C\sm\{0\}$. Each term of this series is an analytic function in
$\C\sm\{0\}$. Hence the sum is analytic in this domain. Summing the
majorants and using the estimate $\int_0^1|\cQ(s)|ds\le{\vk\/|z|}$
we get \er{eN}. $\BBox$

In the following Lemma (proof in Appendix) we will determine
asymptotics of the trace $T$ of the monodromy matrix. Introduce the
auxiliary function $\phi$ which will be used
in this asymptotics:
\[
\lb{phin}
\phi(t,\l)=
\sum_{0\le k<j\le 2}\o^{2(k+j)}e^{i\o^{j}z}e^{i(\o^{k}-\o^{j})zt},
\qqq(t,\l)\in\R\ts\C.
\]

For $ p\ge 1,\a\in \R$  we introduce the real  spaces
$$
 \ell^p=\rt\{f=(f_n)_{n\in \Z},\ \ \| f \|_p<\iy \rt\},
\qqq \ \ \| f \|_p^p=\sum _{n} |f_n|^p <\iy,
$$
$$
\ell_\a^p=\rt\{f=(f_n)_{n\in \Z},\ \ \sum_{n\in
\Z}(1+|2\pi n|^{2\a})|f_n|^p<\iy \rt\}.
$$
We will write $a_n=\ell_\a^p(n)$
iff the sequence $(a_n)_{n\in\Z}\in\ell_\a^p$.

\begin{lemma}
\lb{TrM}
Let $p,q\in L^2(\T)$. Then
the function $T$ satisfies
\[
\lb{aT}
T=\Phi_0+{\Phi_1\/z^2}+{\wt\Phi\/z^3}
\]
where
\[
\lb{P1} \Phi_0=\sum_{k=0}^2e^{iz\o^{k}+{2i\wh p_0\/3\o^{k}z}},\qq
\Phi_1(\l)=-{1\/9}\int_0^1\phi(s,\l)\eta(s)ds,\qq
\eta(s)=\int_{0}^1p(t)p(t-s)dt,
\]
and the function $\wt\Phi$ satisfies
\[
\lb{gSl}
\wt\Phi(\l)=e^{z_0}\ca\qqq O(|z|^{-1})\qq \ \ \as\ \ |\l|\to\iy,
\qqq\arg\l\in[-{\pi\/4},{5\pi\/4}]\\
\ell^1(n)+O(n^{-1})\ \ \as\ \ n\to+\iy,\
\l=-i\bigl({2\pi n\/\sqrt3}\bigr)^3\bigl(1+O(n^{-2})\bigr)\ac\!\!\!\!,
\]
uniformly in $\arg\l\in[-{\pi\/4},{5\pi\/4}]$.
\end{lemma}

\section {Ramifications}
\setcounter{equation}{0}

In this Section we will consider the function $\r$ given by \er{r}.
Now we will prove a Counting Lemma for the ramifications.
Introduce the contours
$$
C_a(r)=\{\l\in\C:|z-a|<r\},\qq a\in\C,\qq r>0,
$$
and the domains
\[
\lb{DomcD}
\cD_{\pm n}=\Bigl\{\l\in\C:\Bigl|z-e^{\pm i{\pi\/6}}{2\pi n\/\sqrt
3}\Bigr| <{\pi\/2\sqrt 3}\Bigr\},\qq\all\qq n\ge 0.
\]

\begin{lemma}
\lb{res}
i) The function $\r$ is entire, real on $\R$, and satisfies
identity \er{rtr} and asymptotics
\[
\lb{asr}
\r=\r_0\bigl(1+O(|z|^{-1})\bigr)\qq \text{as}\ \ |\l|\to\iy,\ \
\l\in\C\sm\cup_{n\in\Z}\cD_n.
\]

ii) For each odd $N>0$ large enough the function $\r$ has
(counting with multiplicities) $2N$ zeros on the domain
$\{|\l|<({N\pi\/\sqrt3})^3\}$ and for each $|n|>{N-1\/2}$ exactly
two zeros in the domain $\cD_n$. There are no other zeros.

iii) The function $\r$ has a finite even number $2m\ge 0$
of real zeros, counted with multiplicities.

\end{lemma}

\no {\bf Proof.}
i)
The standard formula for the discriminant $d$ of the cubic polynomial
$-\t^3+a\t^2-b\t+1$ gives $d=(ab)^2-4(a^3+b^3)+18ab-27$
(see, e.g., \cite{Co}, Ch.7.5)
which implies that $\r$ is entire and
real on $\R$ and satisfies \er{rtr}.

We will show \er{asr}. Let $|\l|\to\iy$.
Asymptotics \er{lom} yields
\[
\lb{t1-3}
{\t_1(\l)-\t_3(\l)\/e^{iz}-e^{i\o^2 z}}
=1+{e^{i(1-\o^2)z}O(z^{-1})+O(z^{-1})\/e^{i(1-\o^2)z}-1}.
\]
We have
$$
-\Re i(1-\o^2)z=\Im(1-\o^2)z ={\sqrt3\/2}(x+\sqrt3y)\ge 0.
$$
Then $|e^{i(1-\o^2)z}|\le 1$ for all $\l\in\C$.
Moreover, using the standard estimate $|\sin z|>{1\/4}e^{|\Im z|}$
as $|z-\pi n|\ge{\pi\/4}$
for all $n\in\Z$ (see \cite{PT}, Lemma 2.1), we deduce that
$$
|e^{i(1-\o^2)z}-1|=2|e^{i(1-\o^2){z\/2}}|\Bigl|\sin{(1-\o^2)z\/2}\Bigl|
>{1\/2}e^{{1\/2}\Re (i(1-\o^2)z)}e^{{1\/2}\Im(1-\o^2)z}={1\/2}
$$
as $\l\in\C\sm\cup_{n\ge 1}\cD_{-n}$. Then asymptotics \er{t1-3} gives
\[
\lb{at1-3}
\t_1(\l)-\t_3(\l)=(e^{iz}-e^{i\o^2 z})
\bigl(1+O(z^{-1})\bigr)\qq\as\qq \l\in\C\sm\cup_{n\ge 1}\cD_{-n}.
\]
The similar arguments show that
\[
\lb{at2-3}
\t_2(\l)-\t_3(\l)=(e^{i\o z}-e^{i\o^2 z})
\bigl(1+O(z^{-1})\bigr)\qq\as\qq  \l\in\C\sm\cup_{n\ge 1}\cD_{-n}.
\]
Furthermore, asymptotics \er{lom} yields
\[
\lb{t1-2}
{\t_1(\l)-\t_2(\l)\/e^{iz}-e^{i\o z}}
=1+{e^{i(1-\o)z}O(z^{-1})+O(z^{-1})\/e^{i(1-\o)z}-1}
=1+{e^{i(\o-1)z}O(z^{-1})+O(z^{-1})\/1-e^{i(\o-1)z}}.
\]
The relations
$$
\Re i(1-\o)z =-\Im (1-\o)z ={\sqrt3\/2}(x-\sqrt3y)
\ca
\ge 0, \Re\l\ge 0\\
<0, \Re\l<0
\ac,
$$
give
$$
|e^{i(1-\o)z}|\le 1,\ \ \all\ \ \l:\Re\l<0;
\qqq\qq |e^{i(\o-1)z}|\le 1,\ \ \all\ \ \l:\Re\l\ge 0.
$$
Moreover, if $\l\in\C\sm\cup_{n\ge 1}\cD_{n}$, then
$$
|e^{i(1-\o)z}-1|=2|e^{i(1-\o){z\/2}}|\Bigl|\sin{(1-\o)z\/2}\Bigl|
>{1\/2}e^{{1\/2}\Re (i(1-\o)z)}e^{{1\/2}\Im(1-\o)z}={1\/2}
$$
as $\Re\l<0$, and similarly $|e^{i(\o-1)z}-1|={1\/2}$ as $\Re\l\ge 0$.
Then asymptotics \er{t1-2} gives
\[
\lb{at1-2}
\t_1(\l)-\t_2(\l)=(e^{iz}-e^{i\o z})
\bigl(1+O(z^{-1})\bigr)\qq\as\qq \l\in\C\sm\cup_{n\ge 1}\cD_{n}.
\]
Substituting asymptotics \er{at1-3}, \er{at2-3}, \er{at1-2} into \er{r} we obtain \er{asr}.

ii) Let $N\ge 1$ be odd and large enough and let $N'>N$ be another
odd. Let $\l$ belong to the contours $C_0({\pi N\/\sqrt3}),
C_0({\pi N'\/\sqrt3})$ and $\pa\cD_n$ for all $
|n|>{N-1\/2}$. Asymptotics \er{asr} yields
$$
|\r(\l)-\r_0(\l)|=|\r_0(\l)|\Bigl|{\r(\l)\/\r_0(\l)}-1\Bigr|
=|\r_0(\l)|O(|z|^{-1})<|\r_0(\l)|
$$
on all contours. Hence, by Rouch\'e's theorem,
$\r(\cdot)$ has as many zeros, as $\r_0(\cdot)$ in each of the
bounded domains and the remaining unbounded domain.
Since $\r_0(\cdot)$ has exactly one zero of multiplicity two
at each $\cD_n,n\in\Z$,
and since $N'>N$ can be chosen arbitrarily large,
the statement i) follows.

iii) Asymptotics \er{asr} shows that $\r(\l)>0$ on the real line for
large $|\l|>0$. Then $\r$ has a finite number of real zeros.
The function $\r$ has even number of zeros in the large disk
due to the statement ii). The
function $\r$ is real on $\R$, hence it has both an even number of
non-real zeros in this disc and an even number of real zeros.
 $\BBox$

Recall that the function $\r$ has a finite number $m$ of real zeros.
Using the results of Lemma \ref{res} we  index the zeros
$r_{n}^\pm,n\in\Z$, of $\r$ by:

\no a) labeling of real zeros:
$$
\begin{aligned}
& m\ \text{even}:\ r_0^-\le r_{-1}^+\le...\le r_{-{m\/2}}^+\le
r_{m\/2}^-\le...\le r_1^-\le r_0^+,
\\
&m\ \text{odd}:\ r_0^-\le r_{-1}^+\le...\le r_{-{m-1\/2}}^-\le
r_{{m-1\/2}}^+\le...\le r_1^-\le r_0^+,
\end{aligned}
$$
\no b) labeling of  non-real zeros:
$$
\begin{aligned}
&\!m\ \text{even}:\
0<\Im r_{{m\/2}}^+\le\Im r_{{m\/2}+1}^-\le\Im r_{{m\/2}+1}^+\le...,
\\
& m\ \text{odd}:\
0<\Im r_{{m+1\/2}}^-\le\Im r_{{m+1\/2}}^+\le...,
\end{aligned}
$$
and
\[
\lb{syr}
r_{-n}^\pm=\ol{r_n^\mp}.
\]

\no {\bf Proof of Theorem \ref{Thr} i), ii).}
i) The statement is proved in Lemma \ref{res} i).

ii) The first identity in \er{sro} was proved in \cite{BK4}. We will
prove the second one. Let  $\t_j=e^{ik_j}, j=1,2,3$, where $\Re
k_j\in(-\pi,\pi]$. We have $k_1+k_2+k_3=0$, since $\t_1\t_2\t_3=1$.
If $k\ne \ell$ and $\l\in\C$ is not a ramification, then $k_j(\l)\ne
k_\ell(\l)$. Identity \er{r} gives
\[
\lb{rqm}
\r=(e^{ik_1}-e^{ik_2})^2
(e^{ik_1}-e^{ik_3})^2
(e^{ik_2}-e^{ik_3})^2
=-64\sin^2{k_1-k_2\/2}
\sin^2{k_1-k_3\/2}
\sin^2{k_2-k_3\/2}.
\]
We have two cases.  If $\l\in\gS_3$, then the first identity in
\er{sro} implies that each $k_j(\l)\in\R, j=1,2,3$, and \er{rqm}
yields $\r(\l)\le 0$.

If $\l\in\s(H)\sm\gS_3$, then exactly one $k_j$, say $k_1$, is real
and $k_3=\ol k_2$ are non-real, since if $\l\in \R$ and $\t_2=e^{i
k_2}$ is a multiplier, then $\t_3=\ol\t_2^{-1}=e^{i\ol k_2}$ is also
a multiplier. Thus identity \er{rqm} implies $\r(\l)=64|\sin{k_1-\ol
k_2\/2}|^2 \sh^2\Im k_2>0$, which yields the second identity in
\er{sro}.

By Lemma \ref{res}, the function $\r$ has a finite even number of
real zeros. Therefore, there exists only a finite number of the
bounded spectral bands with the spectrum of multiplicity $3$.
$\BBox$

We will show in the following Lemma that the large
multipliers in $\C_+$ specify by the dominating multiplier $\t_3$ in
the sense that these multipliers are zeros of the function
\[
\lb{psi}
\p(\l)=\t_3(\l)-(\ol\t_3(\ol\l))^2,
\]
see \er{mir}.
Moreover, here we determine the rough high energy asymptotics
of the ramifications
and the multipliers at the large ramifications.

\begin{lemma}
Let the branches of the multipliers are define by \er{lom}.
Then

i) The multipliers satisfy
\[
\lb{symm}
\ol\t_1^{-1}(\ol\l)=\t_1(\l),\qqq
\ol\t_2^{-1}(\ol\l)=\t_3(\l)
\]
for all $\l\in\L_R$ and for some $R>0$ large enough.

ii) Let $\l=r_{n}^{\pm}$ and let $n\to+\iy$. Then
\[
\lb{amr}
\t_1(\l)=\t_2(\l)
=(-1)^ne^{-{\pi n\/\sqrt3}}\bigl(1+O(n^{-1})\bigr),\qqq
\t_3(\l)=e^{{2\pi n\/\sqrt3}}\bigl(1+O(n^{-1})\bigr),
\]
\[
\lb{amrc}
\ol\t_1(\ol\l)=\ol\t_3(\ol{\l})
=(-1)^ne^{{\pi n\/\sqrt3}}\bigl(1+O(n^{-1})\bigr),\qqq
\ol\t_2(\ol\l)=e^{-{2\pi n\/\sqrt3}}\bigl(1+O(n^{-1})\bigr).
\]

iii) The ramifications satisfy
\[
\lb{ravr}
r_{n}^{\pm}=i\Bigl({2\pi n\/\sqrt3}\Bigr)^3
\bigl(1+O(n^{-2})\bigr)
\qqq\as\qq n\to+\iy.
\]

iv) Let $\l\in\C_+\cap\L_R$ for some $R>0$ large enough, where
the domains $\L_R=\{\l\in\C:|\l|>R\}$. Then
\[
\lb{mir}
\l\ \text{ is a ramification}\qqq \Leftrightarrow\qqq
\p(\l)=0.
\]

\end{lemma}

\no {\bf Proof.}
i) Recall that $\t(\l)$ is a multiplier iff $\ol\t^{-1}(\ol\l)$
is a multiplier. Asymptotics \er{lom} and identity $\ol\o=\o^2$ give
$$
\ol\t_1^{-1}(\ol\l)=e^{iz}\big(1+O(|z|^{-1})\big),\qq
\ol\t_2^{-1}(\ol\l)=e^{i\o^2z}\big(1+O(|z|^{-1})\big),\qq
\ol\t_3^{-1}(\ol\l)=e^{i\o z}\big(1+O(|z|^{-1})\big)
$$
as $|\l|\to\iy$, which yields \er{symm}.

ii), iii)
If $\l=r_{n}^{\pm}$,
then $\t_j(\l)=\t_k(\l)$ for some $j,k=1,2,3$.
Let $n\to+\iy$.
By Lemma \ref{res},
$z=(r_{n}^{\pm})^{1\/3}=e^{i{\pi\/6}}({2\pi n\/\sqrt3}+\d_n)$ where
$|\d_n|<{\pi\/2\sqrt3}$.
Asymptotics \er{lom} gives
$
\t_j(\l)=O(e^{-{\pi n\/\sqrt3}}),j=1,2,
$
and
\[
\lb{t3}
\t_3(\l)=e^{{2\pi n\/\sqrt3}+\d_n}\bigl(1+O(n^{-1})\bigr).
\]
These asymptotics show that $\t_3(\l)\ne \t_j(\l)$ for $j=1,2$
and for all $n\in\N$ large enough.
Then $\t_1(\l)=\t_2(\l)$, which implies the first identity in \er{amr}.

We will prove \er{ravr}.
Let $\l=r_n^\pm$ and let $n\to+\iy$. Then
$$
iz=i(r_{n}^{\pm})^{1\/3}=ie^{i{\pi\/6}}\Bigl({2\pi n\/\sqrt3}+O(n^{-1})\Bigr)
=\o\Bigl({2\pi n\/\sqrt3}+O(n^{-1})\Bigr)
$$
and asymptotics \er{lom} yields
\[
\lb{t21}
\t_1(\l)=e^{\o({2\pi n\/\sqrt3}+\d_n)}\bigl(1+O(n^{-1})\bigr),\qqq
\t_2(\l)=e^{\o^2({2\pi n\/\sqrt3}+\d_n)}\bigl(1+O(n^{-1})\bigr).
\]
Substituting these asymptotics into
the first identity in \er{amr} and using
$\o-\o^2=i\sqrt3$ we obtain
$
e^{i\d_n}=1+O(n^{-1}).
$
Then $\d_n=O(n^{-1})$,
which yields \er{ravr}.

Substituting $\d_n=O(n^{-1})$ into \er{t3}, \er{t21}
we obtain asymptotics \er{amr}.
These asymptotics together with \er{symm} yield \er{amrc}.

iv) Let $\l\in\C_+\cap\L_R$ be a ramification.
Identities \er{symm} and \er{amr} yield
$
\bar\t_3^{-1}(\ol\l)=\t_2(\l)=\t_1(\l).
$
Using the identity
\[
\lb{D13}
\t_1(\l)\t_2(\l)\t_3(\l)=1
\]
we obtain $\p(\l)=0$. Conversely, let $\p(\l)=0$
for some $\l\in\C_+\cap\L_R$. Then identity \er{symm}  implies
$\t_3(\l)=\t_2^{-2}(\l)$. Identity \er{D13} yields
$\t_1(\l)=\t_2(\l)$, therefore $\l$ is a ramification.
$\BBox$

\no {\bf Remark.} 1) We describe the surface of the multipliers for
the case $p=q=0$. We have $\t^0(\l)=e^{i\l^{1\/3}}$ in this case,
then our surface is the 3-sheeted Riemann surface of the function
$\l^{1\/3}$. We have two parametrization of this surface.

First parametrization. We construct the standard (escalator type)
parametrization $\wt\cR^0$ of the surface of the function $\l^{1\/3}$
if we take 3 replicas $\wt\cR_j^0,j=1,2,3$, of the cut plane $\C\sm
i\R_-$
and join the edge of the cut on the sheet $\wt\cR_1$ with the edge of
the cut on the sheet $\wt\cR_2$,
the edge of the cut on the sheet $\wt\cR_2$ with the edge of
the cut on the sheet $\wt\cR_3$,
and the edge of the cut on the sheet $\wt\cR_3$ with the edge of
the cut on the sheet $\wt\cR_1$,
in the usual (crosswise) way, see Fig.~\ref{RiemSur0} a).

Describe the function $\t^0(\l)$.
Consider the sectors $S_j=\o^{j-1}S,j=1,2,3$, on the $z$-plane, where $S$ is given by
\er{S}. In each of the sectors the function $\l^{1\/3}$, and then $\t^0(\l)$,
is univalent. Moreover, for each $j=1,2,3$, the function $\t^0(\l)$ satisfies:
$\t^0(\l)=\t_j^0(\l)=e^{i\o^{j-1}z}$ in the sector $\o^{j-1}S$.
The function  $\l=z^3$ maps each of the sectors $S_j,j=1,2,3$, onto the sheet
$\wt\cR_j^0$ of the
surface $\wt\cR^0$.
For each $j=1,2,3$, the function $\t^0(\l)$ satisfies:
$\t^0(\l)=\t_j^0(\l)$ on the sheet $\wt\cR_j^0$.

Second parametrization. In order to consider below the perturbed
case it will be convenient to use the other parametrization $\cR^0$
of the Riemann surface of the function $\l^{1\/3}$, see
Fig.~\ref{RiemSur0} b). We take $3$ replicas of the cut plane
$\cR_1^0=\C\sm i\R_+,\cR_2^0=\C\sm i\R$ and $\cR_3^0=\C\sm i\R_-$. We
obtain the Riemann surface $\cR^0$ by joining the edges
of the cut $i\R_+$ on $\cR_1^0$ and on $\cR_{2}^0$
and the edges
of the cut $i\R_-$ on $\cR_2^0$
and on $\cR_{3}^0$ in the usual (crosswise) way.

If we deform continuously the surface $\cR^0$ so that the right half-plane
$\Re\l>0$ of the sheet $\cR_1^0$ and the right half-plane
of the sheet $\cR_2^0$ swap places, then we obtain the surface $\wt\cR^0$.
The function $\t^0(\l)$ satisfies:

\no a) $\t^0(\l)=\t_3^0(\l)$ on the sheet $\cR_3^0$;

\no b) $\t^0(\l)=\t_1^0(\l)$ on the left half-plane $\Re\l<0$ of the sheet $\cR_1^0$
and on the right half-plane $\Re\l>0$ of the sheet $\cR_2^0$;

\no c) $\t^0(\l)=\t_2^0(\l)$ on the left half-plane of the sheet $\cR_2^0$
and on the right half-plane of the sheet $\cR_1^0$.

\begin{figure}
\tiny
\unitlength 0.8mm 
\linethickness{0.4pt}
\ifx\plotpoint\undefined\newsavebox{\plotpoint}\fi 
\begin{picture}(92.058,105.804)(8,0)
\multiput(2.586,2.124)(.04185396825,.03372698413){945}{\line(1,0){.04185396825}}
\multiput(3.354,36.876)(.04185396825,.03372698413){945}{\line(1,0){.04185396825}}
\multiput(4.122,71.628)(.04185396825,.03372698413){945}{\line(1,0){.04185396825}}
\put(2.586,1.932){\line(1,0){53.76}}
\put(3.354,36.684){\line(1,0){53.76}}
\put(4.122,71.436){\line(1,0){53.76}}
\multiput(56.154,1.932)(.03665075922,.03373535792){922}{\line(1,0){.03665075922}}
\multiput(56.922,36.684)(.03665075922,.03373535792){922}{\line(1,0){.03665075922}}
\multiput(57.69,71.436)(.03665075922,.03373535792){922}{\line(1,0){.03665075922}}
\put(41.946,33.804){\line(1,0){48.576}}
\put(42.714,68.556){\line(1,0){48.576}}
\put(43.482,103.308){\line(1,0){48.576}}
\multiput(30.81,.972)(.03517557252,.03370992366){1048}{\line(1,0){.03517557252}}
\multiput(31.578,35.724)(.03517557252,.03370992366){1048}{\line(1,0){.03517557252}}
\multiput(32.346,70.476)(.03517557252,.03370992366){1048}{\line(1,0){.03517557252}}
\put(20.058,19.404){\line(1,0){60.288}}
\put(20.826,54.156){\line(1,0){60.288}}
\put(21.594,88.908){\line(1,0){60.288}}
\multiput(30.506,2.228)(.035488152,.033668246){511}{\line(1,0){.035488152}}
\multiput(30.506,36.596)(.035488152,.033668246){511}{\line(1,0){.035488152}}
\multiput(33.3,2.112)(.035488152,.033668246){510}{\line(1,0){.036765957}}
\multiput(33.4,36.48)(.035488152,.033668246){510}{\line(1,0){.036765957}}
\put(48.948,53.246){\line(0,-1){.9737}}
\put(49.085,51.298){\line(0,-1){.9737}}
\put(49.222,49.351){\line(0,-1){.9737}}
\put(49.359,47.403){\line(0,-1){.9737}}
\put(49.496,45.456){\line(0,-1){.9737}}
\put(49.633,43.509){\line(0,-1){.9737}}
\put(49.771,41.561){\line(0,-1){.9737}}
\multiput(49.87,39.492)(.03323077,-.37661538){54}{\line(0,-1){.37661538}}
\put(50.528,53.246){\line(0,-1){.96}}
\put(50.441,51.326){\line(0,-1){.96}}
\put(50.303,49.406){\line(0,-1){.96}}
\put(50.116,47.486){\line(0,-1){.96}}
\put(49.979,45.566){\line(0,-1){.96}}
\put(49.742,43.646){\line(0,-1){.96}}
\put(49.505,41.726){\line(0,-1){.96}}
\multiput(49.538,39.684)(-.0283913,-.42991304){47}{\line(0,-1){.42991304}}
\multiput(30.85,36.8)(.03348837,-.40186047){86}{\line(0,-1){.40186047}}
\multiput(31.85,71.3)(.02848837,-.40186047){86}{\line(0,-1){.40186047}}
\multiput(35.114,71.092)(-.02848837,-.40186047){172}{\line(0,-1){.40186047}}
\multiput(31.926,71.536)(.035628866,.033649485){520}{\line(-1,0){.035628866}}
\multiput(35.234,71.536)(.034553398,.033553398){520}{\line(0,1){.033553398}}
\put(50.128,88.718){\line(0,-1){.9984}}
\put(50.224,86.721){\line(0,-1){.9984}}
\put(50.32,84.724){\line(0,-1){.9984}}
\put(50.416,81.727){\line(0,-1){.9984}}
\put(50.512,80.731){\line(0,-1){.9984}}
\put(50.608,78.734){\line(0,-1){.9984}}
\put(50.704,76.737){\line(0,-1){.9984}}
\put(50.8,74.74){\line(0,-1){.9984}}
\put(50.896,72.743){\line(0,-1){.9984}}
\multiput(51.05,70.628)(.032,-.8106667){20}{\line(0,-1){.8106667}}
\put(52.816,88.718){\line(0,-1){.9691}}
\put(52.778,86.779){\line(0,-1){.9691}}
\put(52.64,84.841){\line(0,-1){.9691}}
\put(52.503,82.903){\line(0,-1){.9691}}
\put(52.465,80.965){\line(0,-1){.9691}}
\put(52.327,79.026){\line(0,-1){.9691}}
\put(52.289,77.088){\line(0,-1){.9691}}
\put(52.152,75.15){\line(0,-1){.9691}}
\put(51.914,73.211){\line(0,-1){.9691}}
\put(51.676,71.273){\line(0,-1){.9691}}
\put(51.439,69.335){\line(0,-1){.9691}}
\multiput(51.398,68.628)(-.152,-2.336){6}{\line(0,-1){2.350}}
\put(79.194,99.084){\makebox(0,0)[cc]{$\wt\cR_1^0$}}
\put(79.194,65.484){\makebox(0,0)[cc]{$\wt\cR_2^0$}}
\put(79.194,30.308){\makebox(0,0)[cc]{$\wt\cR_3^0$}}
\put(69.194,93.084){\makebox(0,0)[cc]{$\t_1^0$}}
\put(69.194,58.084){\makebox(0,0)[cc]{$\t_2^0$}}
\put(69.194,23.308){\makebox(0,0)[cc]{$\t_3^0$}}
\put(80.306,17.356){\makebox(0,0)[cc]{$\Re\l$}}
\put(72.306,37.356){\makebox(0,0)[cc]{$\Im\l$}}
\put(83.306,7.356){\makebox(0,0)[cc]{$a)$}}
\end{picture}
\begin{picture}(92.058,105.804)(-8,0)
\multiput(2.586,2.124)(.04185396825,.03372698413){945}{\line(1,0){.04185396825}}
\multiput(3.354,36.876)(.04185396825,.03372698413){945}{\line(1,0){.04185396825}}
\multiput(4.122,71.628)(.04185396825,.03372698413){945}{\line(1,0){.04185396825}}
\put(2.586,1.932){\line(1,0){53.76}}
\put(3.354,36.684){\line(1,0){53.76}}
\put(4.122,71.436){\line(1,0){53.76}}
\multiput(56.154,1.932)(.03665075922,.03373535792){922}{\line(1,0){.03665075922}}
\multiput(56.922,36.684)(.03665075922,.03373535792){922}{\line(1,0){.03665075922}}
\multiput(57.69,71.436)(.03665075922,.03373535792){922}{\line(1,0){.03665075922}}
\put(41.946,33.804){\line(1,0){48.576}}
\put(42.714,68.556){\line(1,0){48.576}}
\put(43.482,103.308){\line(1,0){48.576}}
\multiput(30.81,.972)(.03517557252,.03370992366){1048}{\line(1,0){.03517557252}}
\multiput(31.578,35.724)(.03517557252,.03370992366){1048}{\line(1,0){.03517557252}}
\multiput(32.346,70.476)(.03517557252,.03370992366){1048}{\line(1,0){.03517557252}}
\put(20.058,19.404){\line(1,0){60.288}}
\put(20.826,54.156){\line(1,0){60.288}}
\put(21.594,88.908){\line(1,0){60.288}}
\multiput(30.506,2.228)(.035488152,.033668246){511}{\line(1,0){.035488152}}
\multiput(30.506,36.596)(.035488152,.033668246){511}{\line(1,0){.035488152}}
\multiput(33.3,2.112)(.035488152,.033668246){510}{\line(1,0){.036765957}}
\multiput(33.4,36.48)(.035488152,.033668246){510}{\line(1,0){.036765957}}
\put(48.948,53.246){\line(0,-1){.9737}}
\put(49.085,51.298){\line(0,-1){.9737}}
\put(49.222,49.351){\line(0,-1){.9737}}
\put(49.359,47.403){\line(0,-1){.9737}}
\put(49.496,45.456){\line(0,-1){.9737}}
\put(49.633,43.509){\line(0,-1){.9737}}
\put(49.771,41.561){\line(0,-1){.9737}}
\multiput(49.87,39.492)(.03323077,-.37661538){54}{\line(0,-1){.37661538}}
\put(51.828,53.246){\line(0,-1){.96}}
\put(51.691,51.326){\line(0,-1){.96}}
\put(51.553,49.406){\line(0,-1){.96}}
\put(51.416,47.486){\line(0,-1){.96}}
\put(51.279,45.566){\line(0,-1){.96}}
\put(51.142,43.646){\line(0,-1){.96}}
\put(51.005,41.726){\line(0,-1){.96}}
\multiput(50.938,39.684)(-.0483913,-.42991304){47}{\line(0,-1){.42991304}}
\multiput(30.85,36.8)(.03348837,-.40186047){86}{\line(0,-1){.40186047}}
\multiput(34.114,36.492)(-.04348837,-.40186047){86}{\line(0,-1){.40186047}}
\multiput(48.726,54.064)(.035628866,.033649485){425}{\line(-1,0){.035628866}}
\multiput(50.226,89.036)(.035628866,.033649485){425}{\line(-1,0){.035628866}}
\multiput(51.934,54.064)(.035053398,.033553398){425}{\line(1,0){.033553398}}
\multiput(52.934,89.036)(.034553398,.033553398){425}{\line(0,1){.033553398}}
\put(50.128,88.718){\line(0,-1){.9984}}
\put(50.224,86.721){\line(0,-1){.9984}}
\put(50.32,84.724){\line(0,-1){.9984}}
\put(50.416,81.727){\line(0,-1){.9984}}
\put(50.512,80.731){\line(0,-1){.9984}}
\put(50.608,78.734){\line(0,-1){.9984}}
\put(50.704,76.737){\line(0,-1){.9984}}
\put(50.8,74.74){\line(0,-1){.9984}}
\put(50.896,72.743){\line(0,-1){.9984}}
\multiput(51.05,70.628)(.032,-.8106667){20}{\line(0,-1){.8106667}}
\put(52.816,88.718){\line(0,-1){.9691}}
\put(52.578,86.779){\line(0,-1){.9691}}
\put(52.34,84.841){\line(0,-1){.9691}}
\put(52.103,82.903){\line(0,-1){.9691}}
\put(51.865,80.965){\line(0,-1){.9691}}
\put(51.627,79.026){\line(0,-1){.9691}}
\put(51.389,77.088){\line(0,-1){.9691}}
\put(51.152,75.15){\line(0,-1){.9691}}
\put(50.914,73.211){\line(0,-1){.9691}}
\put(50.676,71.273){\line(0,-1){.9691}}
\put(50.439,69.335){\line(0,-1){.9691}}
\multiput(50.398,68.628)(-.252,-2.336){6}{\line(0,-1){2.350}}
\put(64.916,103.43){\line(0,-1){.976}}
\put(65.044,101.478){\line(0,-1){.976}}
\put(65.172,99.526){\line(0,-1){.976}}
\put(65.3,97.574){\line(0,-1){.976}}
\put(65.428,95.622){\line(0,-1){.976}}
\put(65.556,93.67){\line(0,-1){.976}}
\put(65.684,91.718){\line(0,-1){.976}}
\put(65.812,89.766){\line(0,-1){.976}}
\put(65.94,87.814){\line(0,-1){.976}}
\put(66.068,85.862){\line(0,-1){.976}}
\put(66.196,83.91){\line(0,-1){.976}}
\put(66.324,81.958){\line(0,-1){.976}}
\put(66.452,80.006){\line(0,-1){.976}}
\multiput(66.714,78.5)(.032,-1.696){6}{\line(0,-1){1.696}}
\put(67.604,103.238){\line(0,-1){.9907}}
\put(67.373,101.256){\line(0,-1){.9907}}
\put(67.143,99.275){\line(0,-1){.9907}}
\put(66.913,97.293){\line(0,-1){.9907}}
\put(66.682,95.312){\line(0,-1){.9907}}
\put(66.452,93.331){\line(0,-1){.9907}}
\put(66.221,91.349){\line(0,-1){.9907}}
\put(65.991,89.368){\line(0,-1){.9907}}
\put(65.761,87.386){\line(0,-1){.9907}}
\put(65.53,85.405){\line(0,-1){.9907}}
\put(65.3,83.423){\line(0,-1){.9907}}
\put(65.069,81.442){\line(0,-1){.9907}}
\put(64.839,79.46){\line(0,-1){.9907}}
\multiput(64.686,77.348)(-.052,-.512){18}{\line(0,-1){.512}}
\put(79.194,99.084){\makebox(0,0)[cc]{$\cR_1^0$}}
\put(79.194,64.484){\makebox(0,0)[cc]{$\cR_2^0$}}
\put(79.194,29.308){\makebox(0,0)[cc]{$\cR_3^0$}}
\put(39.194,91.584){\makebox(0,0)[cc]{$\t_1^0$}}
\put(72.194,93.584){\makebox(0,0)[cc]{$\t_2^0$}}
\put(39.194,56.584){\makebox(0,0)[cc]{$\t_2^0$}}
\put(65.194,49.584){\makebox(0,0)[cc]{$\t_1^0$}}
\put(66.194,15.808){\makebox(0,0)[cc]{$\t_3^0$}}
\put(39.194,21.808){\makebox(0,0)[cc]{$\t_3^0$}}
\put(80.306,17.356){\makebox(0,0)[cc]{$\Re\l$}}
\put(72.306,37.356){\makebox(0,0)[cc]{$\Im\l$}}
\put(83.306,7.356){\makebox(0,0)[cc]{$b)$}}
\end{picture}
\caption{\footnotesize The Riemann surface of the function
$\l^{1\/3}$ , which coincides with the Riemann surface of the multipliers
for the case $p=q=0$, in
a) the standard (escalator type) parametrization,
b) the parametrization convenient for the considering
of the perturbed case}
\lb{RiemSur0}
\end{figure}
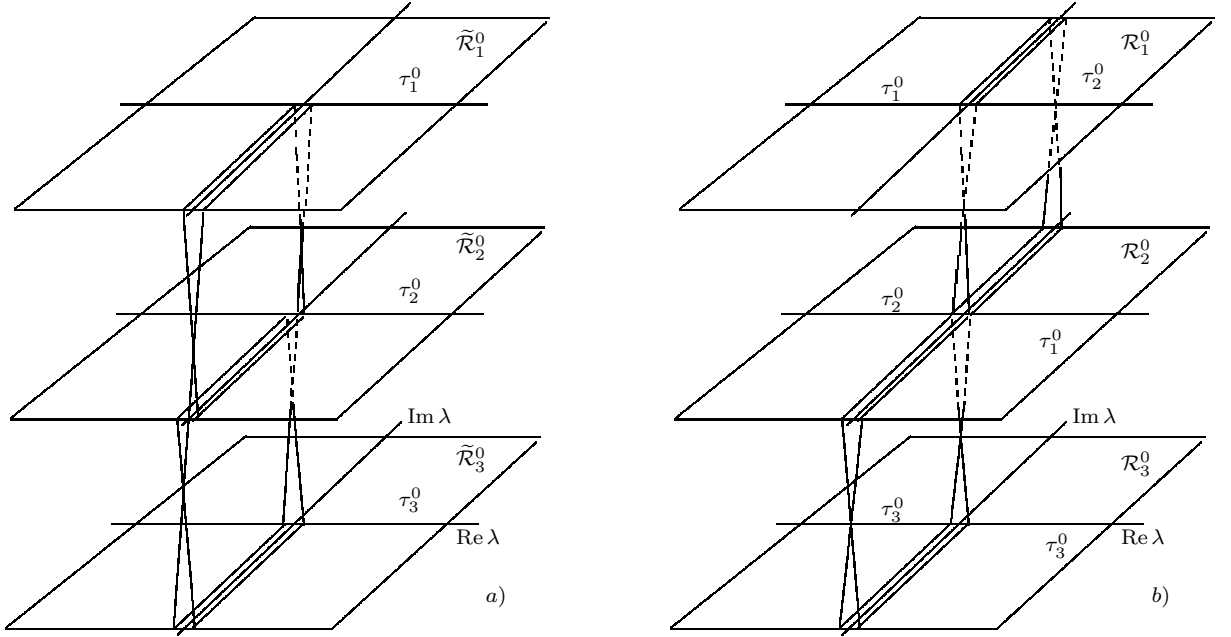

2) Consider the perturbed case.
The  surface $\cR$ of the multipliers for large $|\l|$
is close to the surface $\cR^0$,
see Fig.~\ref{RiemSur}.
Let $R=({N\pi\/\sqrt3})^3$, where $N$ is given by Lemma \ref{res} ii).
Then $r_n^\pm\in\cD_n$ for each $|n|>{N-1\/2}$.
Describe the surface $\cR$ for $|\l|>R$.
We take $3$ replicas $\cR_1,\cR_2$ and $\cR_3$ of the cut plane:
$$
\cR_1\cap\L_R=\L_R\sm\cup_{n\in\N}\G_n,\qq
\cR_2\cap\L_R=\L_R\sm\cup_{n\in\Z}\G_n,\qq
\cR_3\cap\L_R=\L_R\sm\cup_{n\in\N}\G_{-n},
$$
where
$\L_R=\{\l\in\C:|\l|>R\}$,
$
\G_n=[r_{n-1}^+,r_{n}^-]\ss\L_R,n\in\Z,
$
are straight lines.
The surface $\cR$ at large $|\l|$ is obtained
by joining the edges of the cuts $\G_n$ on $\cR_1$ with
the edges of the same cuts on $\cR_{2}$,
the edges of the cuts $\G_{-n}$ on $\cR_2$ with the edges of
the same cuts on $\cR_{3}$ in the crosswise way.

\begin{figure}
\tiny
\unitlength 1mm 
\linethickness{0.4pt}
\ifx\plotpoint\undefined\newsavebox{\plotpoint}\fi 
\begin{picture}(92.058,105.804)(0,0)
\multiput(2.586,2.124)(.04185396825,.03372698413){945}{\line(1,0){.04185396825}}
\multiput(3.354,36.876)(.04185396825,.03372698413){945}{\line(1,0){.04185396825}}
\multiput(4.122,71.628)(.04185396825,.03372698413){945}{\line(1,0){.04185396825}}
\put(2.586,1.932){\line(1,0){53.76}}
\put(3.354,36.684){\line(1,0){53.76}}
\put(4.122,71.436){\line(1,0){53.76}}
\multiput(56.154,1.932)(.03665075922,.03373535792){922}{\line(1,0){.03665075922}}
\multiput(56.922,36.684)(.03665075922,.03373535792){922}{\line(1,0){.03665075922}}
\multiput(57.69,71.436)(.03665075922,.03373535792){922}{\line(1,0){.03665075922}}
\put(41.946,33.804){\line(1,0){48.576}}
\put(42.714,68.556){\line(1,0){48.576}}
\put(43.482,103.308){\line(1,0){48.576}}
\multiput(30.81,.972)(.03517557252,.03370992366){1048}{\line(1,0){.03517557252}}
\multiput(31.578,35.724)(.03517557252,.03370992366){1048}{\line(1,0){.03517557252}}
\multiput(32.346,70.476)(.03517557252,.03370992366){1048}{\line(1,0){.03517557252}}
\put(20.058,19.404){\line(1,0){60.288}}
\put(20.826,54.156){\line(1,0){60.288}}
\put(21.594,88.908){\line(1,0){60.288}}
\multiput(38.106,9.228)(.035488152,.033668246){211}{\line(1,0){.035488152}}
\multiput(38.106,43.596)(.035488152,.033668246){211}{\line(1,0){.035488152}}
\multiput(40.8,9.112)(.035488152,.033668246){210}{\line(1,0){.036765957}}
\multiput(40.8,43.48)(.035488152,.033668246){210}{\line(1,0){.036765957}}
\put(45.402,16.14){\line(1,0){2.496}}
\put(45.402,50.508){\line(1,0){2.496}}
\put(37.914,9.036){\line(1,0){2.688}}
\put(37.914,43.404){\line(1,0){2.688}}
\multiput(30.35,2.24)(.037052632,.033684211){100}{\line(1,0){.037052632}}
\multiput(30.35,36.608)(.037052632,.033684211){100}{\line(1,0){.037052632}}
\multiput(36.986,5.388)(-.033649485,-.033649485){97}{\line(0,-1){.033649485}}
\multiput(36.986,39.756)(-.033649485,-.033649485){97}{\line(0,-1){.033649485}}
\put(33.882,39.948){\line(1,0){3.2}}
\put(33.882,5.58){\line(1,0){3.2}}
\put(52.698,57.228){\line(1,0){2.588}}
\put(53.466,91.98){\line(1,0){2.588}}
\multiput(52.698,57.228)(.034807018,.033684211){171}{\line(1,0){.034807018}}
\multiput(53.466,91.98)(.034807018,.033684211){171}{\line(1,0){.034807018}}
\put(58.342,62.796){\line(1,0){2.304}}
\put(59.61,97.548){\line(1,0){2.304}}
\put(55.194,57.228){\line(1,1){5.76}}
\put(55.962,91.98){\line(1,1){5.76}}
\put(60.962,65.484){\line(1,0){3.00}}
\put(61.53,100.236){\line(1,0){3.004}}
\multiput(64.226,68.564)(-.035628866,-.033649485){90}{\line(-1,0){.035628866}}
\multiput(61.53,100.236)(.035628866,.033649485){90}{\line(-1,0){.035628866}}
\multiput(63.834,65.484)(.033553398,.033553398){90}{\line(1,0){.033553398}}
\multiput(64.602,100.236)(.033553398,.033553398){90}{\line(0,1){.033553398}}
\multiput(30.85,36.8)(.03348837,-.40186047){86}{\line(0,-1){.40186047}}
\multiput(34.114,36.492)(-.04348837,-.40186047){86}{\line(0,-1){.40186047}}
\put(37.844,43.142){\line(0,-1){.9326}}
\put(37.953,41.277){\line(0,-1){.9326}}
\put(38.063,39.411){\line(0,-1){.9326}}
\put(38.173,37.546){\line(0,-1){.9326}}
\put(39.956,43.334){\line(0,-1){.96}}
\put(39.846,41.414){\line(0,-1){.96}}
\put(39.736,39.494){\line(0,-1){.96}}
\put(39.627,37.574){\line(0,-1){.96}}
\put(44.948,50.246){\line(0,-1){.9737}}
\put(45.085,48.298){\line(0,-1){.9737}}
\put(45.222,46.351){\line(0,-1){.9737}}
\put(45.359,44.403){\line(0,-1){.9737}}
\put(45.496,42.456){\line(0,-1){.9737}}
\put(45.633,40.509){\line(0,-1){.9737}}
\put(45.771,38.561){\line(0,-1){.9737}}
\put(47.828,50.246){\line(0,-1){.96}}
\put(47.691,48.326){\line(0,-1){.96}}
\put(47.553,46.406){\line(0,-1){.96}}
\put(47.416,44.486){\line(0,-1){.96}}
\put(47.279,42.566){\line(0,-1){.96}}
\put(47.142,40.646){\line(0,-1){.96}}
\put(47.005,38.726){\line(0,-1){.96}}
\multiput(38.298,36.684)(.03352381,-.4327619){63}{\line(0,-1){.4327619}}
\multiput(39.45,36.876)(-.0333913,-.58852174){47}{\line(0,-1){.58852174}}
\multiput(46.17,36.492)(.03323077,-.37661538){54}{\line(0,-1){.37661538}}
\multiput(46.938,36.684)(-.0333913,-.42991304){47}{\line(0,-1){.42991304}}
\multiput(36.05,36.5)(-.03333333,-.51){60}{\line(0,-1){.51}}
\multiput(34.05,36.5)(.04333333,-.46969697){66}{\line(0,-1){.46969697}}
\put(33.78,39.63){\line(0,-1){.75}}
\put(33.98,38.13){\line(0,-1){.75}}
\multiput(36.68,39.23)(-.033333,-.144444){6}{\line(0,-1){.144444}}
\multiput(36.28,37.496)(-.033333,-.144444){6}{\line(0,-1){.144444}}
\put(53.628,91.718){\line(0,-1){.9984}}
\put(53.724,89.721){\line(0,-1){.9984}}
\put(53.82,87.724){\line(0,-1){.9984}}
\put(53.916,85.727){\line(0,-1){.9984}}
\put(54.012,83.731){\line(0,-1){.9984}}
\put(54.108,81.734){\line(0,-1){.9984}}
\put(54.204,79.737){\line(0,-1){.9984}}
\put(54.3,77.74){\line(0,-1){.9984}}
\put(54.396,75.743){\line(0,-1){.9984}}
\put(54.492,73.747){\line(0,-1){.9984}}
\multiput(54.55,71.628)(.032,-.8106667){18}{\line(0,-1){.8106667}}
\put(56.316,91.718){\line(0,-1){.9691}}
\put(56.078,89.779){\line(0,-1){.9691}}
\put(55.84,87.841){\line(0,-1){.9691}}
\put(55.603,85.903){\line(0,-1){.9691}}
\put(55.365,83.965){\line(0,-1){.9691}}
\put(55.127,82.026){\line(0,-1){.9691}}
\put(54.889,80.088){\line(0,-1){.9691}}
\put(54.652,78.15){\line(0,-1){.9691}}
\put(54.414,76.211){\line(0,-1){.9691}}
\put(54.176,74.273){\line(0,-1){.9691}}
\put(53.939,72.335){\line(0,-1){.9691}}
\multiput(53.898,71.628)(-.202,-2.336){6}{\line(0,-1){2.350}}
\put(58.964,97.478){\line(0,-1){.976}}
\put(59.06,95.526){\line(0,-1){.976}}
\put(59.156,93.574){\line(0,-1){.976}}
\put(59.252,91.622){\line(0,-1){.976}}
\put(59.348,89.67){\line(0,-1){.976}}
\put(59.444,87.718){\line(0,-1){.976}}
\put(59.54,85.766){\line(0,-1){.976}}
\put(59.636,83.814){\line(0,-1){.976}}
\put(59.732,81.862){\line(0,-1){.976}}
\put(59.828,79.91){\line(0,-1){.976}}
\put(59.924,77.958){\line(0,-1){.976}}
\put(60.02,76.006){\line(0,-1){.976}}
\put(60.116,74.054){\line(0,-1){.976}}
\put(60.212,72.102){\line(0,-1){.976}}
\put(60.308,70.15){\line(0,-1){.976}}
\put(60.404,68.198){\line(0,-1){.976}}
\put(60.5,66.246){\line(0,-1){.976}}
\put(60.596,64.294){\line(0,-1){.976}}
\put(61.844,97.286){\line(0,-1){.992}}
\put(61.604,95.302){\line(0,-1){.992}}
\multiput(61.364,93.318)(-.03,-.248){4}{\line(0,-1){.248}}
\put(61.124,91.334){\line(0,-1){.992}}
\put(60.884,89.35){\line(0,-1){.992}}
\multiput(60.644,87.366)(-.03,-.248){4}{\line(0,-1){.248}}
\put(60.404,85.382){\line(0,-1){.992}}
\put(60.164,83.398){\line(0,-1){.992}}
\multiput(59.924,81.414)(-.03,-.248){4}{\line(0,-1){.248}}
\put(59.684,79.43){\line(0,-1){.992}}
\put(59.444,77.446){\line(0,-1){.992}}
\multiput(59.204,75.462)(-.03,-.248){4}{\line(0,-1){.248}}
\put(58.506,65.18){\line(0,1){.992}}
\put(58.266,63.18){\line(0,1){.992}}
\put(61.844,99.974){\line(0,-1){.9767}}
\put(61.977,98.02){\line(0,-1){.9767}}
\put(62.111,96.067){\line(0,-1){.9767}}
\put(62.244,94.114){\line(0,-1){.9767}}
\put(62.378,92.16){\line(0,-1){.9767}}
\put(62.512,90.207){\line(0,-1){.9767}}
\put(62.645,88.253){\line(0,-1){.9767}}
\put(62.779,86.3){\line(0,-1){.9767}}
\put(62.912,84.347){\line(0,-1){.9767}}
\put(63.046,82.393){\line(0,-1){.9767}}
\put(63.179,80.44){\line(0,-1){.9767}}
\put(63.313,78.486){\line(0,-1){.9767}}
\put(63.447,76.532){\line(0,-1){.9767}}
\put(63.581,74.578){\line(0,-1){.9767}}
\put(63.715,72.624){\line(0,-1){.9767}}
\put(63.849,70.67){\line(0,-1){.9767}}
\put(63.983,68.716){\line(0,-1){.9767}}
\put(64.117,66.762){\line(0,-1){.9767}}
\put(64.34,100.166){\line(0,-1){.9677}}
\put(64.171,98.23){\line(0,-1){.9677}}
\put(64.002,96.295){\line(0,-1){.9677}}
\put(63.833,94.36){\line(0,-1){.9677}}
\put(63.664,92.424){\line(0,-1){.9677}}
\put(63.495,90.489){\line(0,-1){.9677}}
\put(63.326,88.554){\line(0,-1){.9677}}
\put(63.157,86.618){\line(0,-1){.9677}}
\put(62.988,84.683){\line(0,-1){.9677}}
\put(62.819,82.747){\line(0,-1){.9677}}
\put(62.65,80.812){\line(0,-1){.9677}}
\put(62.481,78.877){\line(0,-1){.9677}}
\put(62.312,76.941){\line(0,-1){.9677}}
\put(62.143,75.005){\line(0,-1){.9677}}
\put(61.974,73.069){\line(0,-1){.9677}}
\put(61.805,71.133){\line(0,-1){.9677}}
\put(61.636,69.197){\line(0,-1){.9677}}
\put(61.467,67.261){\line(0,-1){.9677}}
\put(62.298,65.325){\line(0,-1){.9677}}
\put(64.916,103.43){\line(0,-1){.976}}
\put(65.044,101.478){\line(0,-1){.976}}
\put(65.172,99.526){\line(0,-1){.976}}
\put(65.3,97.574){\line(0,-1){.976}}
\put(65.428,95.622){\line(0,-1){.976}}
\put(65.556,93.67){\line(0,-1){.976}}
\put(65.684,91.718){\line(0,-1){.976}}
\put(65.812,89.766){\line(0,-1){.976}}
\put(65.94,87.814){\line(0,-1){.976}}
\put(66.068,85.862){\line(0,-1){.976}}
\put(66.196,83.91){\line(0,-1){.976}}
\put(66.324,81.958){\line(0,-1){.976}}
\put(66.452,80.006){\line(0,-1){.976}}
\put(66.58,78.054){\line(0,-1){.976}}
\put(66.708,76.102){\line(0,-1){.976}}
\put(66.836,74.15){\line(0,-1){.976}}
\put(66.964,72.198){\line(0,-1){.976}}
\put(67.092,70.246){\line(0,-1){.976}}
\put(67.604,103.238){\line(0,-1){.9907}}
\put(67.373,101.256){\line(0,-1){.9907}}
\put(67.143,99.275){\line(0,-1){.9907}}
\put(66.913,97.293){\line(0,-1){.9907}}
\put(66.682,95.312){\line(0,-1){.9907}}
\put(66.452,93.331){\line(0,-1){.9907}}
\put(66.221,91.349){\line(0,-1){.9907}}
\put(65.991,89.368){\line(0,-1){.9907}}
\put(65.761,87.386){\line(0,-1){.9907}}
\put(65.53,85.405){\line(0,-1){.9907}}
\put(65.3,83.423){\line(0,-1){.9907}}
\put(65.069,81.442){\line(0,-1){.9907}}
\put(64.839,79.46){\line(0,-1){.9907}}
\put(64.609,77.478){\line(0,-1){.9907}}
\put(64.379,75.496){\line(0,-1){.9907}}
\put(64.149,73.514){\line(0,-1){.9907}}
\put(63.919,71.532){\line(0,-1){.9907}}
\put(63.689,69.55){\line(0,-1){.9907}}
\put(79.194,99.084){\makebox(0,0)[cc]{$\cR_1$}}
\put(79.194,65.484){\makebox(0,0)[cc]{$\cR_2$}}
\put(79.194,31.308){\makebox(0,0)[cc]{$\cR_3$}}
\put(68.21,51.00){\makebox(0,0)[cc]{$r_0^+$}}
\put(58.21,51.00){\makebox(0,0)[cc]{$r_{1}^-$}}
\put(50.322,91.404){\makebox(0,0)[cc]{$r_1^+$}}
\put(54.546,97.356){\makebox(0,0)[cc]{$r_2^-$}}
\put(59.346,101.58){\makebox(0,0)[cc]{$r_2^+$}}
\put(31.21,57.00){\makebox(0,0)[cc]{$r_0^-$}}
\put(41.21,57.00){\makebox(0,0)[cc]{$r_{-1}^+$}}
\put(51.21,15.26){\makebox(0,0)[cc]{$r_{-1}^-$}}
\put(44.298,8.348){\makebox(0,0)[cc]{$r_{-2}^+$}}
\put(40.306,5.356){\makebox(0,0)[cc]{$r_{-2}^-$}}
\put(80.306,17.356){\makebox(0,0)[cc]{$\Re\l$}}
\put(72.306,37.356){\makebox(0,0)[cc]{$\Im\l$}}
\multiput(58.25,88.3)(.0333333,.04){30}{\line(0,1){.04}}
\multiput(58.45,53.5)(.0333333,.04){30}{\line(0,1){.04}}
\put(59.25,89.5){\line(1,0){10.2}}
\put(59.45,54.7){\line(1,0){10.2}}
\multiput(69.45,89.5)(-.03333333,-.03333333){36}{\line(0,-1){.03333333}}
\multiput(69.65,54.7)(-.03333333,-.03333333){36}{\line(0,-1){.03333333}}
\put(58.45,88.1){\line(1,0){9.6}}
\put(58.65,53.3){\line(1,0){9.6}}
\put(67.78,88.03){\line(0,-1){.9714}}
\put(67.894,86.087){\line(0,-1){.9714}}
\put(68.008,84.144){\line(0,-1){.9714}}
\put(68.123,82.201){\line(0,-1){.9714}}
\put(69.18,89.63){\line(0,-1){.975}}
\put(69.13,87.68){\line(0,-1){.975}}
\put(69.08,85.73){\line(0,-1){.975}}
\put(69.03,83.78){\line(0,-1){.975}}
\multiput(68.25,80.9)(.03333333,-.53333333){48}{\line(0,-1){.53333333}}
\multiput(69.25,81.9)(-.0333333,-.9533333){30}{\line(0,-1){.9533333}}
\put(58.18,87.83){\line(0,-1){.96}}
\put(58.286,85.91){\line(0,-1){.96}}
\put(58.393,83.99){\line(0,-1){.96}}
\put(58.5,82.07){\line(0,-1){.96}}
\put(58.606,80.15){\line(0,-1){.96}}
\put(58.713,78.23){\line(0,-1){.96}}
\put(58.82,76.31){\line(0,-1){.96}}
\put(58.926,74.39){\line(0,-1){.96}}
\put(59.38,89.43){\line(0,-1){.9444}}
\put(59.313,87.541){\line(0,-1){.9444}}
\put(59.246,85.652){\line(0,-1){.9444}}
\put(59.18,83.763){\line(0,-1){.9444}}
\put(59.113,81.874){\line(0,-1){.9444}}
\put(59.046,79.985){\line(0,-1){.9444}}
\put(58.98,78.096){\line(0,-1){.9444}}
\put(58.913,76.207){\line(0,-1){.9444}}
\put(58.846,74.319){\line(0,-1){.9444}}
\multiput(58.85,72.9)(.0333333,-.9777778){18}{\line(0,-1){.9777778}}
\multiput(58.85,73.1)(-.0333333,-1.0777778){18}{\line(0,-1){1.0777778}}
\put(31.05,54.7){\line(1,0){9.8}}
\put(31.05,20.1){\line(1,0){9.8}}
\multiput(40.85,54.7)(-.0333333,-.04){30}{\line(0,-1){.04}}
\multiput(40.85,20.1)(-.0333333,-.04){30}{\line(0,-1){.04}}
\put(30.25,53.5){\line(1,0){9.8}}
\put(30.25,18.9){\line(1,0){9.8}}
\multiput(31.05,54.7)(-.0333333,-.05){24}{\line(0,-1){.05}}
\multiput(31.05,20.1)(-.0333333,-.05){24}{\line(0,-1){.05}}
\multiput(31.05,54.7)(-.033333,-.2){6}{\line(0,-1){.2}}
\put(30.18,53.43){\line(0,-1){.9939}}
\put(30.252,51.442){\line(0,-1){.9939}}
\put(30.325,49.454){\line(0,-1){.9939}}
\put(30.398,47.466){\line(0,-1){.9939}}
\put(30.471,45.478){\line(0,-1){.9939}}
\put(30.543,43.49){\line(0,-1){.9939}}
\put(30.616,41.502){\line(0,-1){.9939}}
\put(30.689,39.515){\line(0,-1){.9939}}
\multiput(30.762,37.527)(.01333333,-.53333333){32}{\line(0,-1){1.0777778}}
\put(31.18,54.43){\line(0,-1){.9722}}
\put(31.135,52.485){\line(0,-1){.9722}}
\put(31.091,50.541){\line(0,-1){.9722}}
\put(31.046,48.596){\line(0,-1){.9722}}
\put(31.002,46.652){\line(0,-1){.9722}}
\put(30.957,44.707){\line(0,-1){.9722}}
\put(30.913,42.763){\line(0,-1){.9722}}
\put(30.869,40.819){\line(0,-1){.9722}}
\put(30.824,38.874){\line(0,-1){.9722}}
\put(30.78,36.93){\line(0,-1){.9722}}
\multiput(30.735,34.985)(-.01333333,-.46969697){33}{\line(0,-1){.46969697}}
\put(39.58,53.03){\line(0,-1){.9879}}
\put(39.64,51.054){\line(0,-1){.9879}}
\put(39.701,49.078){\line(0,-1){.9879}}
\put(39.762,47.102){\line(0,-1){.9879}}
\put(39.822,45.127){\line(0,-1){.9879}}
\put(39.883,43.151){\line(0,-1){.9879}}
\put(39.943,41.175){\line(0,-1){.9879}}
\put(40.004,39.199){\line(0,-1){.9879}}
\put(40.065,37.224){\line(0,-1){.9879}}
\put(40.125,35.248){\line(0,-1){.9879}}
\put(40.186,33.272){\line(0,-1){.9879}}
\put(40.246,31.296){\line(0,-1){.9879}}
\put(40.307,29.321){\line(0,-1){.9879}}
\put(40.368,27.345){\line(0,-1){.9879}}
\put(40.428,25.369){\line(0,-1){.9879}}
\put(40.489,23.393){\line(0,-1){.9879}}
\put(40.549,21.418){\line(0,-1){.9879}}
\put(40.78,54.63){\line(0,-1){.9889}}
\put(40.713,52.652){\line(0,-1){.9889}}
\put(40.646,50.674){\line(0,-1){.9889}}
\put(40.58,48.696){\line(0,-1){.9889}}
\put(40.513,46.719){\line(0,-1){.9889}}
\put(40.446,44.741){\line(0,-1){.9889}}
\put(40.38,42.763){\line(0,-1){.9889}}
\put(40.313,40.785){\line(0,-1){.9889}}
\put(40.246,38.807){\line(0,-1){.9889}}
\put(40.18,36.83){\line(0,-1){.9889}}
\put(40.113,34.852){\line(0,-1){.9889}}
\put(40.046,32.874){\line(0,-1){.9889}}
\put(39.98,30.896){\line(0,-1){.9889}}
\put(39.913,28.919){\line(0,-1){.9889}}
\put(39.846,26.941){\line(0,-1){.9889}}
\put(39.78,24.963){\line(0,-1){.9889}}
\put(39.713,22.985){\line(0,-1){.9889}}
\put(39.646,21.007){\line(0,-1){.9889}}
\put(39.194,91.084){\makebox(0,0)[cc]{$\t_1$}}
\put(73.194,93.084){\makebox(0,0)[cc]{$\t_2$}}
\put(36.194,59.084){\makebox(0,0)[cc]{$\t_2$}}
\put(73.194,59.084){\makebox(0,0)[cc]{$\t_1$}}
\put(69.194,21.308){\makebox(0,0)[cc]{$\t_3$}}
\put(26.194,15.308){\makebox(0,0)[cc]{$\t_3$}}
\end{picture}
\caption{\footnotesize The Riemann surface $\cR$ of the multipliers
and the ramifications for the case of the coefficients $p_0+p,q$,
with small $p,q$ and constant $p_0$.}
\lb{RiemSur}
\end{figure}
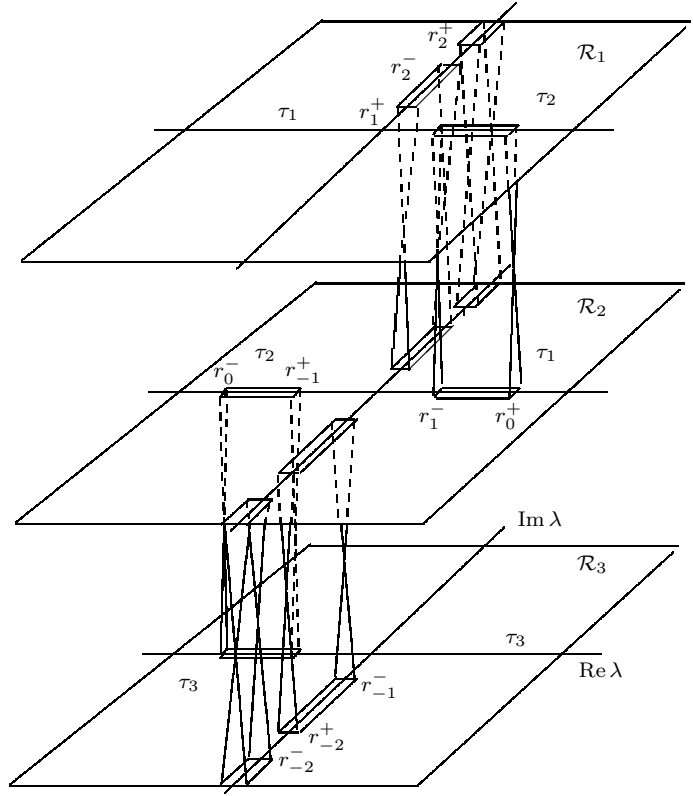

\section{Asymptotics of the ramifications}
\setcounter{equation}{0}

Introduce
the entire function $\xi $ by
\[
\lb{F}
\xi (\l)=4T(\l)-\ol T^2(\ol\l),\qqq\l\in\C.
\]

\begin{lemma}
Let $\l=r_n^\pm$. Then the function $\p(\l)=\t_3(\l)-\ol
\t_3^2(\ol\l)$ satisfies
\[ \lb{tir}
\p(\l)=\Bigl({1\/4}+O(n^{-1})\Bigr) \bigl(\x(\l)+O(e^{-{\pi n\/\sqrt3}})\bigr)
\qq\as\qq n\to+\iy.
\]
\end{lemma}

\no {\bf Proof.}
Identity \er{cM} provides
\[
\lb{D1}
D(\t(\l),\l)
=-\t^2(\l)\bigl(\t(\l)-T(\l)+\t^{-1}(\l)\ol T(\ol\l)
+\t^{-2}(\l)\bigr),
\]
\[
\lb{D2}
\ol D(\t(\ol\l),\ol\l)=-\ol\t(\ol\l)
\bigl(\ol\t^2(\ol\l)-\ol\t(\ol\l)\ol T(\ol\l)+
T(\l)-\ol\t^{-1}(\ol\l)\bigr)
\]
for all $\l\in\C$.
Let $\l=r_{n}^{\pm}$ and let $n\to+\iy$.
Identity \er{D1} and asymptotics \er{amr} imply
$$
D(\t_3(\l),\l)=-\t_3^2(\l)\bigl(\t_3(\l)-T(\l)
+O(e^{-{\pi n\/\sqrt3}})\bigr).
$$
The identity $D(\t_3(\l),\l)=0$ yields
\[
\lb{at4}
\t_3(\l)=T(\l)+O(e^{-{\pi n\/\sqrt3}}).
\]
Identities \er{mir}, \er{D2} and asymptotics \er{amrc}, \er{at4} give
\begin{multline}
\lb{D5}
\ol D(\t_3(\ol\l),\ol\l)
=-\ol \t_3(\ol\l)\bigl(\t_3(\l)-\ol \t_3(\ol\l) \ol T(\ol\l)
+T(\l)+O(e^{-{\pi n\/\sqrt3}})\bigr)
\\
=-\ol \t_3(\ol\l)\bigl(2T(\l)-\ol \t_3(\ol\l) \ol T(\ol\l)
+O(e^{-{\pi n\/\sqrt3}})\bigr).
\end{multline}
Asymptotics \er{amr}, \er{amrc} and identity $T=\t_1+\t_2+\t_3$ give
\[
\lb{aT1}
T(\l)=e^{{2\pi n\/\sqrt3}}\bigl(1+O(n^{-1})\bigr),
\qq
\ol T(\ol\l)=2(-1)^ne^{{\pi n\/\sqrt3}}\bigl(1+O(n^{-1})\bigr).
\]
Asymptotics \er{D5}, \er{aT1} and the identity
$\ol D(\t_3(\ol\l),\ol\l)=0$ yield
\[
\lb{at3}
\ol \t_3^2(\ol\l)
={4T^2(\l)\/\ol T^2(\ol\l)}+O(e^{-{\pi n\/\sqrt3}}).
\]
Asymptotics \er{at4}, \er{at3} give
$$
\p(\l)=\t_3(\l)-\ol \t_3^2(\ol\l)=
T(\l)-{4T^2(\l)\/\ol T^2(\ol\l)}+O(e^{-{\pi n\/\sqrt3}}).
$$
Then asymptotics \er{aT1}
yields \er{tir}.
$\BBox$

Below we will often use the following simple result.

\begin{lemma}
\lb{cf}
Let $f\in C([0,1])$ and let $\a>0,\b\in\R$. Then
\[
\lb{ai}
\int_0^1 e^{(-\a+i\b)x u}f(u)du
={f(0)+o(1)\/(\a-i\b)x}\qq \as\qq x\to+\iy.
\]

\end{lemma}

\no {\bf Proof.}
We have
\[
\lb{ai0}
\int_0^1 e^{(-\a+i\b)x u}f(u)du
={(1+e^{(-\a+i\b) x })f(0)\/(\a-i\b)x}
+\int_0^{1} e^{(-\a+i\b)x u}k(u)du,\qq\all\qq x>0,
\]
where $k(u)=f(u)-f(0).$
Furthermore,
$$
A=\Big|\int_0^{1} e^{(-\a+i\b)x u}k(u)du\Big|\le
\int_0^{\d} e^{-\a x u}|k(u)|du
+\int_{\d}^1 e^{-\a x u}|k(u)|du
$$
$$
\le\max_{[0,\d]}|k(u)|\int_0^{\d} e^{-\a x u}du
+\max_{[0,1]}|k(u)|\int_{\d}^1 e^{-\a x u}du
\le {{\max_{[0,\d]}|k(u)|+ e^{-\a x \d} \max_{[0,1]}|k(u)|\/\a x}}
$$
for any $\d\in(0,1)$. Let  $x\to+\iy$ and let $\d={\log x\/x}$.
Then $A=o(x^{-1})$.
Substituting this estimate into \er{ai0}
we obtain \er{ai}.
$\BBox$

The following Lemma gives the asymptotics of the function $\xi$, given by \er{F},
at the large ramifications in $\C_+$.

\begin{lemma}
Let $p,q\in L^2(\T)$. Let $\l=r_n^\pm=z^3,\z=i\o^2z$, and let $n\to+\iy$. Then
the function $\xi $, given by \er{F}, satisfies
\[
\lb{asF}
\xi (\l)=4e^{\z-{2\wh p_0\/3\z}}\lt[
\sin^2\Bigl({\sqrt3\z\/2}+{\wh p_0\/2\pi n}+O(n^{-3})\Bigr) -{|\wh
p_n|^2\/12(\pi n)^2}+\ell_{3\/2}^1(n)+O(n^{-4})\rt].
\]

\end{lemma}

\no {\bf Proof.}
Asymptotics \er{aT} yields
\[
\lb{aF}
\xi =\xi _0+{\xi _1\/z^2}+{\wt \xi \/z^3}
\]
where
\[
\lb{F012}
\xi _0(\l)=4\Phi_0(\l)-\ol\Phi_0^2(\ol\l),\qqq
\xi _1(\l)=4\Phi_1(\l)-2\ol\Phi_0(\ol\l)\ol\Phi_1(\ol\l),
\]
\[
\lb{A16}
\wt \xi (\l)=4\wt\Phi(\l)-2\ol\Phi_0(\ol\l)\ol{\wt\Phi}
(\ol\l)+O(e^{z_0}n^{-1}).
\]

Asymptotics \er{ravr} yields
$\z={2\pi n\/\sqrt3}+O(n^{-1})$.
Substituting this asymptotics
into identity \er{P1} we get
\[
\lb{Phi0}
\Phi_0(\l)=e^{\z-{2\wh p_0\/3\z}}+O(e^{-{\pi n\/\sqrt3}}),\qq
\ol\Phi_0(\ol\l)=2e^{{\z\/2}-{\wh p_0\/3\z}}\cos\Bigl({\sqrt3 \z\/2}
+{\wh p_0\/\sqrt3\z}\Bigr)+O(e^{- {2\pi n\/\sqrt3}}).
\]
Substituting \er{gSl}, \er{Phi0} into \er{A16} we obtain
\[
\lb{A17}
\wt \xi (\l)
=e^{z_0}\bigl(\ell^1(n)+O(n^{-1})\bigr)
\]
Asymptotics \er{Phi0} give
\[
\lb{aF0}
\xi _0(\l)
=4e^{\z-{2\wh p_0\/3\z}}\sin^2\Bigl({\sqrt3 \z\/2}
+{\wh p_0\/\sqrt3\z}\Bigr)
+O(e^{- {\pi n\/\sqrt3}}).
\]
Assume that
\[
\lb{aF1}
\xi _1(\l)
={4\/9}\o^2e^{\z}\Bigl(|\wh p_n|^2+\ell_{1\/2}^1(n)+O(n^{-2})\Bigr).
\]
Substituting asymptotics \er{aF0}, \er{aF1} into \er{aF} we obtain
$$
\xi (\l)=4e^{\z-{2\wh p_0\/3\z}}\lt(\sin^2
\Bigl({\sqrt3 \z\/2}+{\wh p_0\/\sqrt3\z}\Bigr)
-{|\wh p_n|^2\/9\z^2}+\ell_{3\/2}^1(n)+O(n^{-4})\rt),
$$
which yields \er{asF}.

We will prove \er{aF1}.
Substituting identity \er{P1} into \er{F012} we obtain
\[
\lb{F1pr}
\xi _1(\l)=-{2\/9}\int_0^1\a (u,\l)\eta(u)du
\]
where
\[
\lb{aa}
\qq \eta(u)=\int_0^1p(t)p(t-u)dt,\qq
\a (u,\l)=2\phi(u,\l)-\ol\Phi_0(\ol\l)\ol\phi(u,\ol\l),
\]
$\phi$ is given by \er{phin}.
Identities \er{phin} imply
$$
\phi(t,\l)
=e^{\z}e^{(\o^2-1)\z t}+\o e^{\z}e^{(\o-1)\z t}
+\o^{2}e^{\o^2\z}e^{(\o-\o^2)\z t},\qq\all\qq(t,\l)\in\R\ts\C.
$$
Using $\o-\o^2=i\sqrt3$ we have
$e^{\o^{2}\z}e^{(\o-\o^{2})\z t}=O(e^{-{\pi n\/\sqrt3}})$ and then
\[
\lb{A1}
\phi(t,\l)
=e^{\z}\big(e^{(\o^2-1)\z t}+\o e^{(\o-1)\z t}
+O(e^{-\sqrt3\pi n})\big)
\]
uniformly on $t\in[0,1]$. Moreover,
\[
\lb{A6}
\ol\phi(t,\ol\l)
=e^{-\o^{2}\z}e^{(\o^{2}-1)\z t}+\o e^{-\o\z}e^{(\o-1)\z (1-t)}
+\o^2e^{-\o^{2}\z}e^{(\o^{2}-\o)\z t},\qq\all\qq(t,\l)\in\R\ts\C.
\]
Asymptotics \er{Phi0} implies
\[
\lb{A7}
\ol\Phi_0(\ol\l)=2(-1)^ne^{\z\/2}\bigl(1+O(n^{-1})\bigr).
\]
Relations \er{A6}, \er{A7} yield
\begin{multline}
\lb{A2}
\!\!\!\!\ol\Phi_0(\ol\l)\ol\phi(t,\ol\l)
=2(-1)^ne^{\z}\big(e^{i{\sqrt3\/2}\z}e^{(\o^{2}-1)\z t}+\o e^{-i{\sqrt3\/2}\z}e^{(\o-1)\z (1-t)}
+\o^2e^{i{\sqrt3\/2}\z}e^{(\o^{2}-\o)\z t}\big)\bigl(1+O(n^{-1})\bigr)
\\
=2e^{\z}\big(e^{(\o^{2}-1)\z t}+\o e^{(\o-1)\z (1-t)}
+\o^2e^{(\o^{2}-\o)\z t}\big)\bigl(1+O(n^{-1})\bigr)
\end{multline}
uniformly on $t\in[0,1]$.
Substituting identities \er{A1}, \er{A2} into \er{aa} we obtain
\[
\lb{A3}
\a (t,\l)
=2e^{\z}\Big(\big(\o e^{(\o-1)\z t}-\o e^{(\o-1)\z (1-t)}-\o^2e^{-(\o-\o^{2})\z t}\big)
\bigl(1+O(n^{-1})\bigr)+e^{(\o^2-1)\z t}O(n^{-1})
\Big)
\]
uniformly on $t\in[0,1]$. Using the identity
$
\eta(1-t)=\eta(t)
$
we obtain
\[
\lb{A4}
\int_0^1 e^{(\o-1)\z t}\eta(t)dt
=\int_0^1e^{(\o-1)\z (1-t)}\eta(t)dt.
\]
Substituting \er{A3} into \er{F1pr} and using \er{A4} we obtain
\begin{multline*}
\xi_1(\l)={4\/9}\o^2e^{\z}\int_0^1\Big(e^{(\o^{2}-\o)\z t}
\bigl(1+O(n^{-1})\bigr)+ e^{(\o-1)\z t}O(n^{-1})+e^{(\o^2-1)\z t}O(n^{-1})
\Big)\eta(t)dt
\\
={4\/9}\o^2e^{\z}\int_0^1\Big(e^{-i2\pi n t}
\bigl(1+O(n^{-1})\bigr)+ e^{-(\sqrt3-i)\pi n t}O(n^{-1})+e^{-(\sqrt3+i)\pi n t}O(n^{-1})
\Big)\eta(t)dt.
\end{multline*}
Asymptotics \er{ai} and identity $\int_0^1e^{-i2\pi n t}\eta(t)dt=|\wh p_n|^2$
give \er{aF1}.
$\BBox$

Now, using the results
of the previous Lemmas, we will determine the asymptotics of the ramifications.

\no {\bf Proof of Theorem \ref{Thr} iii).}
Identities $r_{-n}^\pm=\ol{r_n^\mp}$ are proved in \er{syr}.
Let $\l=r_{n}^{\pm}$ and let
$n\to +\iy$. Then  \er{ravr} implies
$\z=i\o^2z={2\pi n\/\sqrt3}+\d_n$ where $\d_n=O(n^{-1})$.
Relations \er{mir}, \er{tir} give $\x(\l)=O(e^{-{\pi n\/\sqrt3}})$.
Substituting \er{asF} into this asymptotics we get
\[
\lb{A71}
\sin^2\Bigl({\sqrt3\d_n\/2}+{\wh p_0\/2\pi n}+O(n^{-3})\Bigr)
={1\/12(\pi n)^2}\Big(|\wh p_n|^2
+\ell_{1\/2}^1(n)+O(n^{-2})\Big).
\]
Recall the estimate $|(w^2+\ve^2)^{1\/2}-w|\le|\ve|$ for all
$w,\ve\in\C$ (see, e.g., \cite{CKP}, Ch.4.5).
Substituting
$$
w=|\wh p_n|,\qq
\ve=\big(\ell_{1\/2}^1(n)+O(n^{-2})\big)^{1\/2}=\ell_{1\/2}^2(n)+O(n^{-1}),
$$
into the last estimate we obtain
$$
\Big(|\wh p_n|^2+\ell_{1\/2}^1(n)+O(n^{-2})\Big)^{1\/2}
=|\wh p_n|+\ell_{1\/2}^2(n)+O(n^{-1}).
$$
Asymptotics \er{A71} gives
$$
\d_n=-{\wh p_0\/\sqrt3\pi n}\pm {1\/3\pi n}\Big(|\wh p_n|
+\ell_{1\/2}^2(n)+O(n^{-1})\Big),
$$
which yields asymptotics \er{are}.
Theorem \ref{Thr} is proved.
$\BBox$


\section{The periodic spectrum}
\setcounter{equation}{0}

In this Section we consider the periodic and
antiperiodic eigenvalues labeling by \er{l2p}.
These eigenvalues are zeros of the entire functions $D(\pm 1,\cdot)$,
where $D$ is given by \er{1c}. Identities \er{cM} give
\[
\lb{iDD}
D(1,\l)=2i\Im T(\l),\qq
D(-1,\l)=2+2\Re T(\l),\qq\text{all}\qq\l\in\R.
\]
If $p=q=0$, then the functions $D(\pm 1,\l)$ have the form
\[
\lb{D0}
D_0(1,\l)=-8i\sin{z\/2}\sin{z\o\/2}\sin{z\o^2\/2},\qq
D_0(-1,\l)=8\cos{z\/2}\cos{z\o\/2}\cos{z\o^2\/2}.
\]
Now we will prove a Counting Lemma for the periodic and antiperiodic eigenvalues.
Introduce the domains
$$
\cK_n=\Big\{\l\in\C:|z-\pi n|<{\pi\/2}\Big\},\qq
\cK_{-n}=\Big\{\l\in\C:|\o z+\pi n|<{\pi\/2}\Big\},\qq
n\in\N.
$$

\begin{lemma}
\lb{CLD}
\lb{pev}
i) The functions $D(\pm 1,\l)$ as $|\l|\to\iy$ satisfy
\[
\lb{asD0}
D(1,\l)=D_0(1,\l)\bigl(1+O(|z|^{-1})\bigr),\qqq\qq\l\in\C\sm\cup_{n\in\Z}\cK_{2n},
\]
\[
\lb{asD1}
D(-1,\l)=D_0(-1,\l)\bigl(1+O(|z|^{-1})\bigr),\qqq\qq\l\in\C\sm\cup_{n\in\Z}\cK_{2n+1}.
\]

ii) For each odd $N>n_0$ for some $n_0\ge 1$ the
function $D(1,\cdot)$ has exactly $N$
zeros, counted with multiplicity,
in the disk $\{\l:|\l|<(\pi N)^3\}$
and for each even $n:|n|>N$
exactly one simple zero in the domain $\cK_{n}$.
There are no other zeros.

iii) For each even $N>n_0$ for some $n_0\ge 1$ the
function  $D(-1,\cdot)$ has exactly  $N$
zeros, counted with multiplicity,
in the disk  $\{\l:|\l|<(\pi N)^3\}$ and for each odd $n:|n|>N$,
exactly one simple zero in the domain $\cK_n$.
There are no other zeros.

\end{lemma}

\no {\bf Proof.} We consider only the function $D(1,\cdot)$.
The proof for $D(-1,\cdot)$ is similar.

i) Identities \er{iDD} and estimates \er{bk42} imply
\[
\lb{eD1}
|D(1,\l)-D_0(1,\l)|\le2|T(\l)-T_0(\l)|
\le {6\vk\/|z|}e^{z_0+\vk},\qq\text{all}\qq|\l|\ge 1.
\]
Substituting the estimate
$|\sin w|>{1\/4}e^{|\Im w|}$ as $|w-\pi n|\ge{\pi\/4}$
for all $n\in\Z$ into \er{D0} and using
the relations
$$
|\Im z|+|\Im z\o|+|\Im z\o^2|=|y|+{|\sqrt3 x-y|\/2}+{\sqrt3 x+y\/2}\ge\sqrt3 x+y=2z_0
$$
we obtain
\[
\lb{eD2}
|D_0(1,\l)|=8\Bigl|\sin{z\/2}\Bigr|\Bigl|\sin{z\o\/2}\Bigr|\Bigl|\sin{z\o^2\/2}\Bigr|
>{1\/8}e^{{1\/2}(|\Im z|+|\Im z\o|+|\Im z\o^2|)}\ge{e^{z_0}\/8}
\]
for all $\l\in\C\sm\cup_{n\in\Z}\cK_{2n}$.
Estimates \er{eD1} and \er{eD2} yield \er{asD0}.

ii), iii)
Let $N\ge 1$ be odd and large enough and let $N'>N$ be another odd.
Let $\l$ belong to the
contours $C_0(\pi N),C_0(\pi N'),C_{\pi n}({\pi\/2}),
|n|>N$, $n$ is even.
Asymptotics \er{asD0} yield
$$
|D(1,\l)-D_0(1,\l)|=|D_0(1,\l)|\Bigl|{D(1,\l)\/D_0(1,\l)}-1\Bigr|
={D_0(1,\l)O(1)\/|z|}<|D_0(1,\l)|
$$
on all contours.
Hence, by Rouch\'e's theorem, $D(1,\cdot)$ has as many zeros,
as $D_0(1,\cdot)$ in each of the
bounded domains and the remaining unbounded domain. Since
$D_0(1,\cdot)$ has exactly one simple zero
at each $\l_{n}^0=(\pi n)^3,n\in\Z$,
and since $N'>N$ can be chosen arbitrarily large,
the statement for $D(1,\l)$ follows.
$\BBox$

The following Lemma shows that the periodic and antiperiodic eigenvalues
$\l_n$ at high energy are zeros of the function $\t_1^2-1$.
The other multipliers remote from the unit circle at large real $\l$.
Nevertheless, these eigenvalues can specify by the multiplier
$\t_3$ as well as the ramifications.
Moreover, we will show that the function ${T(\l)\/\ol T(\ol\l)}-(-1)^n$
is exponentially close to zero at the large eigenvalue $\l_n$,
and determine the rough high energy asymptotics of these eigenvalues.

\begin{lemma}
Let $n\to+\iy$.
Then the periodic and antiperiodic eigenvalues satisfy
\[
\lb{t3p}
(-1)^n=\t_1(\l_n)={\t_3(\l_n)\/\ol\t_3(\l_n)}
={T(\l_n)\/\ol T(\l_n)}\bigl(1+O(e^{-{\sqrt3\/2}\pi n})\bigr),
\]
\[
\lb{rae}
\l_n=(\pi n)^{3}\bigl(1+O(n^{-2})\bigr).
\]
\end{lemma}

\no {\bf Proof.}
Let $\l=\l_n,n\to+\iy$. Lemma \ref{CLD} gives
$z=\pi n+\d_n\in\R$ and $|\d_n|<{\pi\/2}$.
Asymptotics \er{lom} gives
\[
\lb{at1r}
\t_3(\l)
=O(e^{{\sqrt3\/2}\pi n}).
\]
Then $|\t_3(\l)|\ne 1$
and identity  \er{symm} implies $\t_2(\l)=\ol\t_3^{-1}(\l)$. Then
$$
(-1)^n=\t_1(\l)={1\/\t_2(\l)\t_3(\l)}={\ol\t_3(\l)\/\t_3(\l)},
$$
which yields the first identities in \er{t3p}.

Estimates \er{bk42} yield
\[
\lb{at6r}
T(\l)=T_0(\l)(1+O(z^{-1}))=(e^{iz}+e^{i\o z}+e^{i\o^2 z})(1+O(z^{-1}))
=O(e^{{\sqrt3\/2}\pi n}).
\]
Substituting \er{at1r} and \er{at6r}
into \er{cM} we obtain
$$
D(\t_3(\l),\l)
=\t_3^2(\l)\bigl(-\t_3(\l)+T(\l)+O(1)\bigr).
$$
The identity $D(\t_3(\l),\l)=0$ and \er{at6r}
imply
$$
\t_3(\l)=T(\l)+O(1)=T(\l)\bigl(1+O(e^{-{\sqrt3\/2}\pi n})\bigr),
$$
which gives the last asymptotics in \er{t3p}.

Asymptotics \er{lom}
implies $(-1)^n=\t_1(\l)=e^{iz}(1+O(n^{-1}))$.
Substituting $z=\pi n+\d_n$ into this asymptotics we obtain
$e^{i\d_n}=1+O(n^{-1})$. Then $\d_n=O(n^{-1})$ and $z=\pi n+O(n^{-1})$
which yields \er{rae}.
$\BBox$

\no {\bf Proof of Theorem \ref{Per}  i)}.
Let $\l=\l_{n},n\to+\iy$.
Asymptotics \er{rae} shows that $z=\pi n+\d_n\in\R,\d_n=O(n^{-1})$.
Asymptotics \er{aT} yields
\[
\lb{Tr}
T(\l)=e^{iz\o^{2}
+{2i\wh p_0\/3\o^{2}z}}\bigl(1+O(e^{-{\sqrt3\/2}\pi n})\bigr)
+{\Phi_1(\l)\/z^2}+{O(e^{z_0})\/n^4}.
\]
Identity \er{phin} gives
$$
\phi(t,\l)=e^{iz\o^2}\!\bigl(e^{i(\o-\o^{2})zt}+\o e^{i(1-\o^{2})zt}
+\o^2 e^{i(1-\o)(1+t)z}\bigr)
=e^{iz\o^2}\!\bigl(
e^{-\sqrt3zt}+\o e^{-\sqrt3e^{-i{\pi\/6}}zt}
+O(e^{-{\sqrt3\/2}\pi n})
\bigr)
$$
uniformly on $t\in[0,1]$.
Substituting this asymptotics into \er{P1} and using \er{ai} we obtain
$$
\Phi_1(\l)=-{e^{iz\o^2}\/9}\Bigl(\int_0^1\bigl(e^{-\sqrt3zu}
+\o e^{-\sqrt3e^{-i{\pi\/6}}zu}
\bigr)\eta(u)du+O(e^{-{\sqrt3\/2}\pi n})\Bigr)=e^{iz\o^2}o(n^{-1}).
$$
Substituting the last asymptotics into \er{Tr} we get
$$
T(\l)=e^{iz\o^{2}+{2i\wh p_0\/3\o^{2}z}}\bigl(1+o(n^{-3})\bigr),\qq
\ol T(\l)=e^{-iz\o-{2i\wh p_0\/3\o z}}\bigl(1+o(n^{-3})\bigr).
$$
Substituting these asymptotics into \er{t3p}
we obtain
$
e^{-iz-{2i\wh p_0\/3z}}
=(-1)^n+o(n^{-3}),
$
which gives
$$
e^{-i\d_n-{2i\wh p_0\/3(\pi n+\d_n)}}
=1+o(n^{-3}).
$$
Then
$$
\d_n=-{2\wh p_0\/3(\pi n+\d_n)}+o(n^{-3})
=-{2\wh p_0\/3\pi n}-{4\wh p_0^2\/9(\pi n)^3}+o(n^{-3}),
$$
which implies \er{apa} for $n\to+\iy$.
The asymptotics for $n\to-\iy$ can be obtained by substituting $-t$
instead of $t$ in equation \er{1b}.
$\BBox$

We need the following Hadamard factorizations
of the functions $D(\pm 1,\cdot)$.

\begin{lemma}
\lb{CCP}
 The functions $D(\pm1,\l)$ satisfy
\[
\lb{HFD}
D(1,\l)=i(\l_0-\l)\prod_{n\in\Z\sm\{0\}}{\l_{2n}-\l\/\l_{2n}^0},
\]
\[
\lb{HFD-}
D(-1,\l)=8\prod_{n}{\l_{2n-1}-\l\/\l_{2n-1}^0},
\]
uniformly on any bounded subset of $\C$, where $\l_n^0=(\pi n)^3,n\in\Z$.
\end{lemma}

\no {\bf Proof.}
We consider the functions $D(1,\cdot)$. The proof for $D(-1,\cdot)$
is similar.
Asymptotics \er{apa} show that
the infinite product in \er{HFD}
converges  uniformly on any bounded subset of $\C$
to the entire function of $\l$,
whose zeros are precisely $\l_{2n},n\in\Z$.
Identity \er{D0} yields
\[
\lb{D11}
D_0(1,\l)=-i\l\prod_{n\ne 0}{\l_{2n}^0-\l\/\l_{2n}^0}.
\]
Asymptotics \er{asD0} and identity \er{D0} show
that the entire function $D(1,\l)$ has order 1.
Moreover, its zeros have asymptotics \er{apa}.
Then
\[
\lb{HF3}
D(1,\l)=ie^{A\l+B}(\l_0-\l)\prod_{n\ne 0}{\l_{2n}-\l\/\l_{2n}^0},
\qq\text{for some}\qq A,B\in\C.
\]
Identities \er{D11}, \er{HF3} yield
\[
\lb{lDD}
\log{D(1,\l)\/D_0(1,\l)}
=A\l+B+\log{\l-\l_0\/\l}+
\sum_{n\ne 0}\log{\l_{2n}-\l\/\l_{2n}^0-\l}.
\]
Let $\l\to+i\iy$. Asymptotics \er{asD0} implies
$\log{D(1,\l)\/D_0(1,\l)}=O(|z|^{-1})$. Moreover, we have
$$
\sum_{n\ne 0}\log{\l_{2n}-\l\/\l_{2n}^0-\l}
=\sum_{n\ne 0}\log\lt(1+{\l_{2n}-\l_{2n}^0\/\l_{2n}^0-\l}\rt)
=\sum_{n\ne 0}\log\lt(1+{O(n)\/\l_{2n}^0-\l}\rt)
=o(1).
$$
Identity \er{lDD} gives $A=B=0$.
Identity \er{HF3} gives \er{HFD}.
$\BBox$

Now we will prove the results about the recovering of the function $\r$
and the spectrum $\s(H)$ by the periodic (or antiperiodic) spectrum.

\no {\bf Proof of Theorem \ref{Per} ii)}.
Let $\{\l_{n},n\ \text{even}\}$
be the periodic spectrum and $\l_*$ be the antiperiodic eigenvalue.
Identities \er{HFD}, \er{iDD} give the function $D(1,\cdot)$
and $\Im T(\l)=-{i\/2}D(1,\l)$.
Then we reconstruct $\Re T(\l)$ up to some constant.
This constant, and then the function $T$,
can be determined from the identity
$\Re T(\l_*)={1\/2}D(-1,\l_*)-1=-1$.
Identity \er{rtr} provides the discriminant $\r$.
Identity \er{sro} gives the spectrum $\gS_3$ of the multiplicity three.

Using the similar calculations we recover the function $T$
and the spectrum of the multiplicity three
by the antiperiodic spectrum and one periodic eigenvalue.
$\BBox$

\section{Small coefficients}
\setcounter{equation}{0}

We prove Theorem \ref{1.3}. Here we use some methods developed for
fourth order operators  with  the small 1-periodic coefficients
\cite{BK1}, \cite{BK2}. We consider the equation
\[
\lb{1be}
i y'''+ \ve (ipy'+i(py)'+q y)=\l y,\qqq \l\in\C.
\]
In this case the matrix-valued function $M(t,\l)=M(t,\l,\ve),
(t,\l,\ve)\in\R\ts\C\ts\R,$ given by \er{deM} is a solution of
equation
\[
\lb{me1e} M'-P(\l)M=\ve Q(t)M,\qq M(0,\l,\ve)=\1_3
\]
where  the $3\ts 3$ matrices $P$ and $Q$ are given by \er{mu}.

Consider the case $\ve =0$.
The matrix-valued function $M_0(t,\l)=M(t,\l,0)$
has the form
$
M_0
=e^{tP}. $ Each function $M_0(t,\cdot), t\in\R$, is entire.
Eigenvalues of the matrix $M_0$ have the form $e^{izt},e^{i\o
zt},e^{i\o^2 zt}$, since eigenvalues of the matrix $P$ are given by
$iz,i\o z,i\o^2 z$. Estimates $|e^{iz\o^jt}|\le e^{z_0|t|}$ imply
\[
\lb{eu1}
|M_0(t,\l)|\le e^{z_0|t|},\qqq\text{all}\qq
(t,\l)\in\R\ts\C,
\]
where $z_0$ is given by \er{kz0}
and a matrix norm is given by \er{mnorm}.

Consider the case $\ve \ne 0$.
The solution $M(t,\l,\ve)$ of problem \er{me1e}
satisfies the integral equation
\[
\lb{iev}
M(t,\l,\ve)=M_0(t,\l)+\ve\int_0^tM_0(t-s,\l)Q(s)M(s,\l,\ve)ds.
\]
The standard iterations in \er{iev} lead to the standard series
\[
\lb{evj}
M(t,\l,\ve)=\sum_{n\ge 0}\ve^nM_n(t,\l),\qq
M_n(t,\l)=\int_0^tM_0(t-s,\l)Q(s)M_{n-1}(s,\l)ds,\qq n\ge 1.
\]

We need the uniform estimates of the monodromy matrix $M(1,\l,\ve)$.

\begin{lemma}
\lb{T21}
For each $t\in\R$ the series \er{evj} converges absolutely
and uniformly on any bounded subset of
$\C^2$.
Each matrix-valued function $M(t,\cdot,\cdot), t\in [0,1]$ is
entire in $(\l,\ve)\in \C^2$ and satisfies:
\[
\lb{2if}
|M(1,\l,\ve)|\le e^{z_0+\vk},\qq
|M(1,\l,\ve)-\sum_{n=0}^{N-1}M_n(1,\l,\ve)|\le
|\ve|^N\vk^Ne^{z_0+|\ve|\vk}
\]
for all $(N,\l,\ve)\in\N\ts\C^2$ where $\vk=\|p\|+\|q\|$.

\end{lemma}

\no {\bf Proof.}  Consider the case $t\ge 0$. The proof
for $t<0$ is similar. Identity \er{evj} gives
\[
\lb{2ig} M_n(t,\l)=\int\limits_{0< t_1<...<
t_n<t_{n+1}=t}\prod\limits_{k=1}^{n}
\Big(M_0(t_{k+1}-t_k,\l)Q(t_k)\Big)M_0(t_1,\l)dt_1dt_2...dt_n,
\]
the factors are ordering from right to left.
Substituting estimates \er{eu1} into identities
\er{2ig} we obtain
\[
\lb{eM1} |M_n(t,\l)|\le{e^{z_0t}\/n!}\Big(\int_0^t|Q(s)|ds\Big)^n,
\qqq \all \qq (n,t,\l)\in\N\ts\R_+\ts\C.
\]
These estimates show that for each fixed $t\ge0 $ the formal series
\er{evj} converges absolutely and uniformly on any bounded subset of
$\C^2$. Each term of this series is an entire function of
$(\l,\ve)$. Hence the sum is an entire function. Summing the
majorants and using the estimate $\int_0^1|Q(s)|ds\le\vk$
we obtain \er{2if}.
$\BBox$

We introduce the {\it discriminant}
$\r(\l,\ve),(\l,\ve)\in\C^2$, of the polynomial
$\det(M(\l,\ve)-\t \1_3)$ by identity \er{r}.

\begin{lemma}\lb{Thr1}
i) The function $\r(\l,\ve)$ is entire in $\C^2$ and satisfies:
\[
\lb{ars} \r(\l,\ve)=\r_0(\l)\bigl(1+O(\ve)\bigr) \qq\as\qq\ve\to 0,
\]
uniformly in $\l$ on any bounded subset of
$\cD=\C\sm\cup_{n\in\Z}\cD_n,\cD_n$ are given by \er{DomcD}.

ii)  Let  $|\ve|<c$ for some $c>0$ small enough. Then the function
$\r(\cdot,\ve )$ has exactly two zeros, counted with multiplicities,
in each domain $\cD_n,n\in\Z$. There are no
other zeros.
In particular, the function $\r(\cdot,\ve)$ has no any real zeros in
the domain $|\l|\ge 1$.

\end{lemma}

\no {\bf Proof.}
i) Identity \er{rtr} and Lemma \ref{T21} show that the function
$\r(\l,\ve)$ is entire. Estimates \er{2if} give
$M(1,\l,\ve)=M_0(1,\l)+O(\ve)$ as $\ve\to 0$ uniformly in $\l$ on
any compact in $\C$. Let $\l\in\cD$. Then  all eigenvalues
$e^{iz},e^{i\o z},e^{i\o^2 z}$ of the matrix $M_0(1,\l)$ are simple
and the standard matrix perturbation theory (see, e.g., \cite{HJ}, Corollary 6.3.4) gives
that the eigenvalues $\t_j(\l,\ve),j=1,2,3$, of the matrix
$M(\l,\ve)$ satisfy
$$
\t_j(\l,\ve)=e^{i\o^{j-1}z}+O(\ve)
=e^{i\o^{j-1}z}\bigl(1+O(\ve)\bigr)\qq\as\qq \ve\to 0
$$
uniformly in $\l$ on any bounded subset of $\cD$.
Substituting these asymptotics into \er{r}
and using the identity
$$
\r_0=(e^{iz}-e^{i\o z})^2(e^{iz}-e^{i\o^2 z})^2(e^{i\o z}-e^{i\o^2 z})^2
$$
we obtain \er{ars}.

ii) Using asymptotics \er{ars} and repeating the arguments
from the proof of Lemma \ref{res} ii) we obtain
the statement.
$\BBox$

Introduce the entire functions
$T_n(\l)=\Tr M_n(1,\l),n\ge 0.$
Estimates \er{2if} imply
\[
\lb{e2}
T(\l,\ve)=\Tr M(1,\l,\ve)
=T_0(\l)+\ve T_1(\l)+\ve^2 T_2(\l)+\ve^3 T_3(\l)+O(\ve^4)
\qq\as\qq\ve\to 0
\]
uniformly in $\l$ on any compact in $\C$.
Below we will use the following relations.

\begin{lemma}
Let $p\in L^1(\T)$ satisfy $\int_0^1 p(t)dt=0$.
Then the functions $T_1$ and $T_2$ satisfy
\[
\lb{T0e}
T_1=0,
\]
\[
\lb{T1e}
\Re T_2(\l)=-3h\bigl(1+O(\l)\bigr)\qq\as\qq\l\to 0
\]
where $h$ is given by \er{F1}.
\end{lemma}

\no {\bf Proof.}  Identity \er{evj} implies
$$
T_1=\Tr M_1(1,\cdot)=\Tr\int_0^1 e^{(1-t)P}Q(t)e^{tP}ds =\Tr e^{P}
\int_0^1 Q(t)dt=0,
$$
which yields \er{T0e}.
Moreover,
\[
\lb{T2} T_2=\Tr M_2(1,\cdot) =\Tr\int_0^1dt\int_0^t
e^{(1-t+s)P}Q(t)e^{(t-s)P}Q(s)ds.
\]
We rewrite identities \er{mu} for
the matrices $P,Q$ in the form
$$
P=P_0-i\l P_1,\qq Q=-pP_0^*+i qP_1,\qq P_0=\ma 0&1&0\\0 &0&1\\
0&0&0\am,\qq
P_1=\ma 0&0&0\\0 &0&0\\ 1&0&0\am.
$$
Using
the identities
$$
e^{tP_0}=\1_3+tP_0+{t^2\/2}P_0^2=
\ma1& t& {t^2\/2}\\
0&1&t\\
0&0&1\am,
$$
we obtain
$$
e^{tP(\l)}Q(s)=e^{tP_0}Q(s)(\1_3+O(\l))
=(-p(s)K_1(t)+iq(s)K_2(t))(\1_3+O(\l))
$$
as $\l\to 0$ uniformly on $(t,s)\in[0,1]^2$ where
$$
K_1=e^{tP_0}P_0^*=\ma t& {t^2\/2}& 0\\
1&t&0\\
0&1&0\am,\qq K_2=e^{tP_0}P_1=\ma {t^2\/2}&0& 0\\
t&0&0\\
1&0&0\am.
$$
Using the identities
$$
\Tr K_1(1-u) K_1(u)=2(1-u)u+{(1-u)^2\/2}+{u^2\/2} ={1\/2}+u(1-u),
$$
$$
\Tr K_2(1-u) K_2(u)={(1-u)^2u^2\/4},
$$
we obtain
$$
\Re\Tr e^{(1-t+s)P}Q(t)e^{(t-s)P}Q(s)=J(t,s)(1+O(\l))
$$
as $\l\to 0$ uniformly on $(t,s)\in[0,1]^2$ where
$$
J(t,s)=p(t)p(s)\Bigl({1\/2}+u(1-u)\Bigr) -q(t)q(s){(1-u)^2u^2\/4},
\qq u=t-s.
$$
Substituting this asymptotics into \er{T2} we obtain
$$
\Re T_2(\l) =\int_0^1dt\int_0^t J(t,s)ds(1+O(\l))
={1\/2}\int_0^1dt\int_{t-1}^t J(t,s) ds(1+O(\l))
$$
as $\l\to 0$, since $J(t,s)=J(s,t-1)$. Using  the simple identity
$$
\int_0^1dt\int_{t-1}^tg(t-s)f(t)f(s)ds =\sum_{k\in\Z}|\wh
f_k|^2\wh g_k
$$
for all $f\in L^1(0,1), g,g'\in L^2(0,1), g(0)=g(1)=0$, and
identities
$$
\int_0^1u(1-u)e^{-i2\pi nu}du=-{1\/2(\pi n)^2},\qqq
\int_0^1u^2(1-u)^2e^{-i2\pi nu}du=-{3\/2(\pi n)^4}
$$
for all $n\ne 0$, we obtain \er{T1e}.
$\BBox$

Now we will determine the asymptotics of the first spectral interval
of multiplicity 3 for the small coefficients.

\no {\bf Proof of Theorem \ref{1.3}.}  Identities \er{t01} imply
\[
\lb{e1}
T_0(\l)=e^{iz}+e^{i\o z}+e^{i\o^2z}=3-{i\l\/2}-{\l^2\/240}+O(\l^3)\qq\as\qq\l\to 0.
\]
Recall that the functions $T(\l,\ve),\r(\l,\ve)$ are entire in $(\l,\ve)\in\C^2$.
Let the entire functions $a(\l,\ve),b(\l,\ve)$
and the numbers $b_j,j=2,3$, be given by
$$
T(\l,\ve)=3+a(\l,\ve)+ib(\l,\ve),\qq b(\l,\ve)=\Im T(\l,\ve),\qq\all\qq
(\l,\ve)\in\R^2,\qq b_j=\Im T_j(0).
$$
Substituting relations \er{T0e}, \er{T1e}, \er{e1} into \er{e2} we obtain
\[
\lb{sp1} T(\l,\ve)=3-{i\l\/2}-{\l^2\/240}
-\bigl(3h-ib_2\bigr)\ve^2+\ve^3T_3(0)+O(\l^3)+O(\l\ve^2)+O(\ve^4)
\]
as $(\l,\ve)\to(0,0)$.
Asymptotics \er{sp1} gives
\[
\lb{sp5}
a(\l,\ve)=-3h\ve^2+O(\l^2)+O(\l\ve^2)+O(\ve^3),\qq
b(\l,\ve)=\m(\l,\ve)+O(\l^3)+O(\l\ve^2)+O(\ve^4)
\]
as $(\l,\ve)\to (0,0)$ where
\[
\lb{e4} \m=-{\l\/2}+{r\/2},\qqq r(\ve)=2b_2\ve^2+2b_3\ve^3.
\]
Identity \er{rtr} gives
\[
\lb{sp4} \r=a^3(a+4)+b^2\bigl(108+2(a+18)a+b^2\bigr).
\]
Substituting asymptotics \er{sp5} into \er{sp4} we obtain
$$
\r(\l,\ve)=f(\m,\ve)
=108\Bigl(\m^2-h^3\ve^6+O(\m^4)+O(\m^2\ve^2)+O(\m\ve^4)+O(\ve^7)\Bigr)
$$
as $(\m,\ve)\to(0,0)$  where
\[
\lb{sc2}
\l=r-2\m
\]
and $f(\m,\ve)$ is entire in $(\m,\ve)$. Introduce the
new variable $u$ by $\m=u\ve^3$. Then
\[
\lb{D12}
f(u\ve^3,\ve)=108\ve^6 E(u,\ve),\qqq E(u,\ve) =u^2-h^3 +O(\ve)
\]
as $\ve\to 0$ uniformly on any compact in
$\C$, the function $E(u,\ve)$ is entire in $(u,\ve)$.

Consider the equation $E(u,\ve)=0$. Let $h\ne 0$. Using ${\pa\/\pa u}E(\pm
h^{3\/2},0)\ne 0$ and the Implicit Function Theorem we deduce that
there exist two real analytic functions $u^\pm(\ve)$ in the disk
$\{|\ve|<c\}$ for some $c>0$ such that each $u^{\pm}(\ve)$ is a
zero of the function $E(\cdot,\ve)$. Asymptotics \er{D12} yields
$$
u^\pm(\ve)=\pm h^{3\/2}+O(\ve)\qq\as\qq\ve\to 0.
$$
Substituting $\mu=u^\pm(\ve)\ve^3$ into the identity \er{sc2}
we deduce that there exist two
real analytic functions $r^\pm(\ve)$ in the disk $\{|\ve|<c\}$ such
that
$$
r^\pm(\ve)=r(\ve)\pm 2h^{3\/2}\ve^3+O(\ve^4)\qq\as\qq\ve\to 0
$$
which yields \er{l0}.

Consider $h>0$. Let $\ve>0$ be small enough. Then
$r^-(\ve)<r(\ve)<r^+(\ve)$. Asymptotics \er{sc2}
shows that $\r(r(\ve),\ve)=f(0,\ve)<0$. Then $\r(\cdot,\ve)<0$
on the whole interval $(r^-(\ve),r^+(\ve))$ and, by Lemma \ref{Thr1}
iii), $\r(\cdot,\ve)\ge 0$ out of this interval. Then, due to
\er{sro}, the spectrum of $H_\ve$ in this interval has multiplicity
3 and the other spectrum has multiplicity 1. The proof for the case
$\ve<0$ is similar.

If $h<0$, then, by Lemma \ref{Thr1} iii), the function $\r$
has no any real zeros and $\r(\cdot,\ve)>0$ on the whole real axis.
Hence all the spectrum has multiplicity 1. $\BBox$

\section {Appendix}
\setcounter{equation}{0}

In this Section we will prove Lemma \ref{TrM}.
Introduce the functions $e_{jk}(t,\l)$ by
\[
\lb{ejk}
e_{jk}(t,\l)=e^{i\o^{j-1}z} e^{i(\o^{k-1}-\o^{j-1})zt},
\qqq\all\qq(t,\l)\in\R\ts\C,\qq j,k=1,2,3.
\]

\begin{lemma}
\lb{Asa}
Let $f,g\in L^2(\T)$.
Then all functions $\a_{jk},1\le j<k\le 3$, given by
\[
\lb{ia}
\a_{jk}(\l)=\int_0^1e_{jk}(t,\l)\int_0^1f(u)g(u-t)dudt,
\]
satisfy asymptotics
\[
\lb{gSl1}
\a_{jk}(\l)=e^{z_0}\ca\qqq O(|z|^{-1})\qq \ \ \as\ \ |\l|\to\iy,
\qqq\arg\l\in[-{\pi\/4},{5\pi\/4}]\\
\ell^1(n)+O(n^{-1})\ \as\ \ n\to+\iy,\
\l=-i\bigl({2\pi n\/\sqrt3}\bigr)^3\bigl(1+O(n^{-2})\bigr)\ac\!\!\!\!,
\]
uniformly in $\arg\l\in[-{\pi\/4},{5\pi\/4}]$.
\end{lemma}

\no {\bf Proof.}
Let $\l=-i({2\pi n\/\sqrt3})^3(1+O(n^{-2}))$ as $n\to+\iy$.
Then $z=i{2\pi n\/\sqrt3}+O(n^{-1})$ or $z=e^{-i{\pi\/6}}{2\pi n\/\sqrt3}+O(n^{-1})$.
Consider the first case, the proof for the second one is similar.
Using the identities
\[
\begin{aligned}
\lb{eejk1}
e^{-i\o^2z}e_{12}(t,\l)=e^{i(\o-\o^2)z}e^{i(1-\o)zt}=e^{-{\sqrt 3\/2}(2 -(1+\sqrt3i)t)z},
\\
e^{-i\o^2z}e_{13}(t,\l)=e^{i(1-\o^{2})zt}=e^{-{\sqrt 3-3i\/2}zt},
\qq
e^{-i\o^2z}e_{23}(t,\l)=e^{i(\o-\o^{2})zt}=e^{-\sqrt 3zt}
\end{aligned}
\]
for all $(t,\l)\in\R\ts\C$,
we obtain
\begin{multline*}
e^{-i\o^2z}e_{12}(t,\l)=e^{-(\sqrt3-i)\pi nt}
(1+O(n^{-1})),
\qq
e^{-i\o^2z}e_{13}(t,\l)=e^{-(\sqrt 3+i)\pi nt}
(1+O(n^{-1})),
\\
e^{-i\o^2z}e_{23}(t,\l)=e^{-i2\pi nt}(1+O(n^{-1}))
\qq\as\qq n\to+\iy\qq\text{uniformly on}\  t\in[0,1].
\end{multline*}
Substituting this asymptotics into identity \er{ia}
and using \er{ai} we obtain that
the function $a_{jk}$ satisfies asymptotics \er{gSl1} for
$\l=-i({2\pi n\/\sqrt3})^3(1+O(n^{-2}))$.

Identities \er{eejk1} yield
$$
|e^{-i\o^2z}e_{12}(t,\l)|
=e^{-\sqrt 3(x(1-t)+\xi t)},
\qq
|e^{-i\o^2z}e_{13}(t,\l)|
=e^{-\sqrt 3\xi t},
\qq
|e^{-i\o^2z}e_{23}(t,\l)|
=e^{-\sqrt 3xt}
$$
for all $(t,\l)\in[0,1]\ts\C$, where $\xi={x+y\sqrt3\/2}$.
Substituting these estimates into identity \er{ia} we obtain
\[
\lb{ePh21}
|e^{-i\o^2 z}\a_{jk}(\l)|\le
\max_{t\in[0,1]}\Big|\int_0^1f(u)g(u-t)du\Big|\int_0^1
\bigl(e^{-\sqrt 3(x(1-t)+\xi t)}
+e^{-\sqrt 3\xi t}+e^{-\sqrt 3xt}\bigr)dt
\]
for all $\l\in\C$.
Let $\arg\l\in[-{\pi\/4},{5\pi\/4}]$.
Then $\arg z\in[-{\pi\/12},{5\pi\/12}]$
and
$
\max\{x,\xi\} \ge|z|\sin{\pi\/12}.
$
Estimates \er{ePh21} yield \er{gSl1}.
$\BBox$

\begin{lemma}
Let $p,q\in L^2(\T)$. Then
the functions
$T_k=\Tr\cM_k(1,\cdot),k\in\N$, satisfy
\[
\lb{T0}
T_1=i{2\wh p_0\theta_2\/3z},\qqq
T_2=-{2\wh p_0^2\theta_1\/9z^2}+{\Phi_1\/z^2}+{\wt T_2\/z^{3}},\qqq
T_3=-i{4\wh p_0^3\theta_0\/81z^3}+{\wt T_3\/z^{3}}
\]
where $\Phi_1$ is given by \er{P1} and
$$
\theta_m(\l)=\sum_{j=0}^2\o^{mj}e^{i\o^j z},\qq m=0,1,2,\qq\l\in\C,
$$
the functions $\wt T_2,\wt T_3$ satisfy \er{gSl1}, and
\[
\lb{Tk}
T_k(\l)=e^{z_0}O(|z|^{-4})\qq\as\qq|\l|\to\iy,\qq\all\qq k\ge 4.
\]

\end{lemma}

\no {\bf Proof.}
Estimates \er{eN} give \er{Tk}. Identities \er{js1}, \er{cK12}
and $\int_0^1q(t)dt=0$ give
\[
\lb{iQjj}
\int_0^1\cQ_{jj}(s,\l)ds=i{2\/3z}\wh p_0\o^{2(j-1)}.
\]
Then identities
$$
T_1(\l)=\Tr\int_0^1e^{iz(1-s)\O}\cQ(s,\l)e^{izs\O}ds
=\Tr e^{iz\O}\int_0^1\cQ(s,\l)ds=\sum_{j=1}^3e^{iz\o^{j-1}}\int_0^1\cQ_{jj}(s,\l)ds,
$$
yield the first identity in \er{T0}.
We will prove the second one. We have
\[
\lb{T20}
T_2(\l)=\Tr\int_0^1\int_0^t e^{iz(1-t)\O}\cQ(t,\l)e^{iz(t-s)\O}
\cQ(s,\l)e^{izs\O}dsdt
=\sum_{j,k=1}^3a_{jk}(\l)
\]
where
\[
\lb{ajk}
a_{jk}(\l)=\int_0^1\int_0^te_{jk}(t-s,\l)v_{jk}(t,s,\l)dsdt,
\qq v_{jk}(t,s,\l)=\cQ_{jk}(t,\l)\cQ_{kj}(s,\l),
\]
and $e_{jk}$ has the form \er{ejk}. Identity \er{iQjj} give
$$
a_{jj}(\l)
={e^{iz\o^{j-1}}\/2}\Bigl(\int_0^1\cQ_{jj}(t,\l)dt\Bigr)^2
=-{2\wh p_0^2\o^{j-1}e^{iz\o^{j-1}}\/9z^2},\qq\all\qq j=1,2,3.
$$
Substituting these identities  into \er{T20} we obtain
\[
\lb{T2pr}
T_2(\l)=-{2\wh p_0^2\/9z^2}\theta_1(\l)+
\sum_{j,k=1\atop j\ne k}^3a_{jk}(\l).
\]
Identity \er{ejk} yields
$
e_{kj}(t,\l)=e_{jk}(1-t,\l)$ for all $j,k=1,2,3,(t,\l)\in\R\ts\C.$
Then \er{ajk} gives
$$
a_{kj}(\l)=\int_0^1du\int_0^ue_{jk}(1-u+s,\l)v_{jk}(s,u,\l)ds
=\int_0^1dt\int_{t-1}^0e_{jk}(t-u,\l)v_{jk}(t,u,\l)du,
$$
which yields
$$
a_{jk}(\l)+a_{kj}(\l)=\int_0^1dt\int_{t-1}^t
e_{jk}(t-s,\l)v_{jk}(t,s,\l)ds
=\int_0^1e_{jk}(u,\l)\int_{0}^1v_{jk}(t,t-u,\l)dtdu
$$
for all $j,k=1,2,3$, $\l\in\C$.
Identities \er{js1}, \er{cK12} give
\[
\lb{T21pr}
\sum_{j,k=1\atop j\ne k}^3a_{jk}(\l)
=\sum_{1\le j<k\le 3}\int_0^1due_{jk}(u,\l)
\int_{0}^1v_{jk}(t,t-u,\l)dt
={\Phi_1(\l)\/z^2}+{\wt\Phi_1(\l)\/z^3}
+O\Big({e^{z_0}\/|z|^4}\Big)
\]
as $|\l|\to\iy$
where
$$
\wt\Phi_1(\l)=
-{1\/9}\sum_{1\le j<k\le 3}\int_0^1e_{jk}(u,\l)
\int_{0}^1\bigl(p(t)q(t-u)+p(t-u)q(t)\bigr)dtdu.
$$
By Lemma \ref{Asa},
the function $\wt\Phi_1$ satisfies \er{gSl1}.
Substituting \er{T21pr} into \er{T2pr}
we obtain the second
identity in \er{T0} and asymptotics \er{gSl1} for $\wt T_2$.

We will prove identity \er{T0} for $T_3$. We have
\begin{multline*}
T_3(\l)=\Tr\int\limits_0^1\int\limits_0^t\int\limits_0^s e^{iz(1-t)\O}
\cQ(t,\l)e^{iz(t-s)\O}\cQ(s,\l)
e^{iz(s-u)\O}\cQ(u,\l)e^{izu\O}dudsdt
\\
=\sum_{j,k,\ell=1}^3
\int_0^1\int_0^t\int_0^s
e^{iz(\o^{j-1}(1+u-t)+\o^{k-1}(t-s)+\o^{\ell-1}(s-u))}\cQ_{jk}(t,\l)
\cQ_{k\ell}(s,\l)\cQ_{\ell j}(u,\l)dudsdt.
\end{multline*}
Identity \er{iQjj} and identity \er{js1} for $\cQ$ give
\[
\lb{T3pr}
T_3(\l)=-{4i\wh p_0^3\/81\l}\theta_0+{i\/27\l}\big(-2B_1(\l)+B_2(\l)\big)
+O\Big({e^{z_0}\/|z|^4}\Big)
\]
as $|\l|\to\iy$
where
\[
\lb{bjkl}
B_1=\sum_{j,k=1\atop j\ne k}^3\o^{j+2k}\b_{jk},\qq\b_{jk}=b_{jjk}+b_{jkj}+b_{kjj},
\qqq B_2=\sum_{j,k=1\atop j\ne k}^3b_{jk\g},
\]
\[
\lb{bjkll}
b_{jk\ell}=\int_0^1\int_0^t\int_0^s
e^{iz(\o^{j-1}(1+u-t)+\o^{k-1}(t-s)+\o^{\ell-1}(s-u))}
f(t,s,u)dudsdt,\qq f(t,s,u)=p(t)p(s)p(u),
\]
$\g=\g(j,k)\in\{1,2,3\},\g\ne j,\g\ne k$.
Assume that the functions $B_1, B_2$ satisfy asymptotics \er{gSl1}.
Then identity \er{T3pr} gives the third identity in \er{T0}
and asymptotics \er{gSl1} for $\wt T_3$.

We will prove that the functions $B_1, B_2$
satisfy asymptotics \er{gSl1}.
Consider the function $B_1.$
We have
$$
b_{jjk}(\l)=\int_0^1\int_0^t\int_0^se_{jk}(s-u,\l)f(t,s,u)dudsdt
=\int_0^1\int_0^t\int_{t-1}^0e_{jk}(t-s,\l)f(t,s,u)dudsdt,
$$
$$
b_{jkj}(\l)=\int_0^1\int_0^t\int_0^se_{jk}(t-s,\l)f(t,s,u)dudsdt,
$$
$$
b_{kjj}(\l)=\int_0^1\int_0^t\int_0^se_{jk}(1-t+u,\l)f(t,s,u)dudsdt
=\int_0^1\int_{t-1}^0\int_{t-1}^se_{jk}(t-s,\l)f(t,s,u)dudsdt.
$$
Substituting these identities into \er{bjkl} we obtain
$$
\b_{jk}(\l)
=\int\limits_0^1\int\limits_{t-1}^t\int\limits_{s}^t e_{jk}(t-s,\l)
f(t,s,u)dudsdt
=\int\limits_0^1e_{jk}(v,\l)\int\limits_{0}^1p(t-v)p(t)
\big(\wt p(t)-\wt p(t-v)\big)dtdv
$$
where $\wt p(t)=\int_0^tp(u)du.$
By Lemma \ref{Asa}, the functions $\b_{jk}$,
and then the function $B_1$, satisfy asymptotics \er{gSl1}.

Consider the function $B_2.$
Identities \er{bjkll} yield
\[
\lb{A15}
b_{jk\g}(\l)
=\int_0^1dt\int_0^t\wt e_{jk}(t,s,\l)h(t,s)ds,
\]
$$
b_{kj\g}(\l)
=\int_0^1dt\int_0^tds\ \wt e_{jk}(t,s,\l)\int_{s}^tp(u-s)p(u-t)p(u)du
$$
for all $\l\in\C,j,1\le j<k\le 3$ where
$$
h(t,s)=\int_t^1p(u-t+s)p(u-t)p(u)du,
\qq
\wt e_{jk}(t,s,\l)=e_{jk}(t,\l)e^{i(\o^{\g-1}-\o^{k-1})zs},
$$
$e_{jk}$ are given by \er{ejk}.
Consider the function $b_{123}$,
the estimates for the other functions $b_{jk\g},j,k=1,2,3,j\ne k$, are similar.
We have
\[
\lb{ph3}
e^{-i\o^2z}\wt e_{12}(t,s,\l)=e^{i(\o-\o^2)z}e^{iz(t-s-\o t+\o^{2}s)}
=e^{-{\sqrt3\/2}z(2-t-s-i\sqrt3(t-s))},
\]
which gives
$
|e^{-i\o^2 z}\wt e_{12}(t,s,\l)|\le
e^{-\sqrt3x(1-t)}
$
for all $(t,s,\l)\in[0,1]^2\ts\C, s\le t$.
Substituting this estimate into \er{A15}
and using the continuity of the function $h(t,s)$
in $s$ we obtain
$$
|e^{-i\o^2 z}b_{123}(\l)|\le\int_0^1\!\!dt\!\int_0^te^{-\sqrt3x(1-t)}|h(t,s)|ds
\le\int_0^1\max_{s\in[0,t]}|h(t,s)|\int_0^te^{-\sqrt3x(1-t)}dsdt
=O\Big({1\/|z|}\Big)
$$
as $|\l|\to\iy$ uniformly in $\arg\l\in[-{\pi\/4},{5\pi\/4}]$.
Thus the function $b_{123}$ satisfy \er{gSl1} for
$\arg\l\in[-{\pi\/4},{5\pi\/4}]$.

Let $\l=-i({2\pi n\/\sqrt3})^3(1+O(n^{-2}))$ as $n\to+\iy$.
Then $z=i{2\pi n\/\sqrt3}+O(n^{-1})$
or $z=e^{-i{\pi\/6}}{2\pi n\/\sqrt3}+O(n^{-1})$.
Consider the first case, the proof for the second one is similar.
Identities \er{ph3} give
$$
e^{-i\o^2 z}\wt e_{12}(t,s,\l)=e^{-\sqrt3\pi n(t-s)}
\bigl(e^{i\pi n(t+s)}+O(n^{-1})\bigr).
$$
Substituting this asymptotics into \er{A15} and using \er{ai}
we obtain
\begin{multline*}
|e^{-i\o^2 z}b_{123}(\l)|
\le\int_0^1dt\int_0^te^{-\sqrt3\pi n(t-s)}|h(t,s)|ds\bigl(1+O(n^{-1})\bigr)
\\
=\int_0^1due^{-\sqrt3\pi n u}\int_u^1|h(t,t-u)|dt\bigl(1+O(n^{-1})\bigr)=O(n^{-1}),
\end{multline*}
which shows that the functions
$b_{123}$ satisfy asymptotics \er{gSl1}.

The similar arguments show that all functions $b_{jk\g}$,
and then the function $B_2$,
satisfy asymptotics \er{gSl1}, which proves Lemma.
$\BBox$

\no {\bf Proof of Lemma \ref{TrM}.}
Substituting \er{t01}, \er{T0}, \er{Tk} into the
identity $T=\sum_{n\ge 0}T_n$ we obtain
\er{aT}.
$\BBox$

\no\small {\bf Acknowledgments.}  This work was supported by the
Ministry of education and science of the Russian Federation, state
contract 14.740.11.0581.

\end{document}